\documentclass[journal]{rmaa}

\usepackage{paralist}

\usepackage{psfrag,color}

\usepackage[latin1]{inputenc}
\usepackage{textcomp}
\usepackage{multirow}

\title{Testing {an Entropy Estimator related to} the Dynamical State of Galaxy Clusters} 

\author{
  J. M. Z\'u\~niga,\altaffilmark{1} 
  C. A. Caretta,\altaffilmark{1}
  A. P. Gonz\'alez,\altaffilmark{2} 
  and E. Garc\'ia-Manzan\'arez\altaffilmark{1}}

\altaffiltext{1}{Departamento de Astronom\'\i{}a,
  Universidad de Guanajuato, M\'exico.}
\altaffiltext{2}{Departamento de Matem\'aticas, Universidad del Cauca, Colombia}

\shortauthor{Z\'u\~niga et al.}
\shorttitle{$H_{Z}$ Cluster Entropy Estimator}

\fulladdresses{
\item Johan M. Z\'u\~niga: Departamento de Astronom\'\i{}a, Universidad de Guanajuato, Apartado Postal 36000, Guanajuato, Gto., M\'exico (jm.zuniga@ugto.mx).}

\listofauthors{J. M. Z\'u\~niga, A. Collaborator, \& L. Author}
\indexauthor{Z\'u\~niga, J. M.}
\indexauthor{Collaborator, A.}
\indexauthor{Author, L.}

\abstract{We propose the entropy estimator $H_Z$, calculated from global dynamical parameters, in an attempt to capture the degree of evolution of galaxy systems. We assume that the observed (spatial and velocity) distributions of member galaxies in these systems evolve over time towards states of higher dynamical relaxation (higher entropy), becoming more random and homogeneous in virial equilibrium. Thus, the $H_Z$-entropy should correspond to the gravitacional assembly state of the systems. This was probed in a sample of 70 well sampled clusters in the Local Universe whose gravitational assembly state, classified from optical and X-ray analysis of substructures, shows clear statistical correlation with $H_Z$. This estimator was also tested on a sample of clusters (halos) from the IllustrisTNG simulations, obtaining results in agreement with the observational ones.}

\resumen{Proponemos el estimador de entrop\'ia $H_Z$, calculado a partir de par\'ametros din\'amicos globales, en un intento de capturar el grado de evoluci\'on de los sistemas de galaxias. Asumimos que las distribuciones (espaciales y de velocidad) observadas de galaxias miembros en estos sistemas evolucionan en el tiempo hacia estados de mayor relajaci\'on din\'amica (mayor entrop\'ia), volvi\'endose m\'as aleatorias y homog\'eneas en el equilibrio virial. As\'i, la entrop\'ia $H_Z$ deber\'ia corresponder con el estado de ensamblaje gravitacional de los sistemas. Esto se prob\'o en una muestra de 70 c\'umulos bien muestreados del Universo Local cuyo estado de ensamblaje gravitacional, clasificado a partir del an\'alisis \'optico y de rayos X de las subestructuras, muestra una clara correlaci\'on estadistica con $H_Z$. Este estimador tambi\'en fue testeado sobre una muestra de c\'umulos (halos) de las simulaciones IllustrisTNG, obteniendo resultados acordes con los obtenidos para los c\'umulos observados.}

\addkeyword{galaxies: clusters: general}
\addkeyword{galaxies: statistics}
\addkeyword{galaxies: kinematics and dynamics}
\addkeyword{gravitation}
\addkeyword{equation of state}

\begin{document}
% Typeset article header
\maketitle

%--------- --------- --------- --------- --------- --------- ---------
%%%%%%%%%%%%%%%%%%%%%%%%%%%%%%%%%%%%%%%%%%%%%%%%%%%%%%%%%
% Section 1
%%%%%%%%%%%%%%%%%%%%%%%%%%%%%%%%%%%%%%%%%%%%%%%%%%%%%%%%%
\section{Introduction}
\label{sec:intro}

Galaxies are not uniformly distributed in the Universe, instead they undergo gravitational clustering that leads to an intricate three-dimensional structure in the shape of a network of knots, filaments, walls and voids \citep[\textit{e.g.},][]{lib2018}. This cosmic web, revealed in both the observed distribution of galaxies \citep[\textit{e.g.},][]{GeH89,ir2020} and in cosmological simulations \citep[\textit{e.g.},][]{Dav85,sp2005},
is called the Large-Scale Structure of the Universe \citep[LSS, \textit{e.g.},][]{pe1980,ei2010}, and contains systems of galaxies at different scales embedded in it \citep[\textit{e.g.},][]{ei1984,cau2014}. Such systems go from small groups, like the Local Group that hosts the Milky Way, to superclusters, the largest and youngest coherent structures formed under gravitational influence in the Universe \citep[reaching up to $\sim 100$ Mpc long, \textit{e.g.},][]{oo1983,Boh21}.

Knot-shaped regions, that is, quasi-spherical overdensities of galaxies with radii ranging from $\sim 1.0$ to $\sim 5.0$ Mpc, 
are commonly known as galaxy groups or clusters. If we define {these galaxy systems} richness in terms of galaxies with masses of the order of the Milky Way or greater, it is usually considered that groups have between 3 and 30 galaxies, poor clusters have between 30 and 50 galaxies, and rich clusters have more than 50 galaxies, counting them over a projected sky area of radius $R_A\sim 2.1$ Mpc according to the criteria first proposed by \citet{ab1958} \citep[see also,][]{ab1989,ba1996}. However, this classification is only an \textit{ad hoc} criterion since there is actually a continuum in the richness of these systems and their physical limits are uncertain. A more elegant and physically motivated way to {characterize} clusters is to define them based on their expected virialized zones, that is, as galaxy regions where the mean density is $\Delta_c$ times the critical density of the Universe at the considered redshift \citep[typical values are $\Delta_c \sim100-200$, \textit{e.g.},][]{car1997b,bry1998,tl2015}, which provides better estimates of their mean dimensions and dynamical properties in the cosmology used. 

{The baryonic ---stellar and gaseous--- matter contained in galaxies and in the form of intracluster} {X-ray emitting hot gas \citep[intracluster medium, ICM, \textit{e.g.},][]{boh2010}} {is estimated \citep[\textit{e.g.},][]{wt1992,lim2005} to account for only 15\% of the total mass of a galaxy system, while the remaining 85\% are provided by the dark matter (DM).} {Nevertheless,} in a first approximation, {one can consider a group/cluster} as a collisionless {ensemble of galaxies} moving in the mean gravitational field generated by its total mass \citep[\textit{e.g.},][]{sc2015}. In this context, galaxies {may be taken as} fundamental observational tracers \citep[or primary units, \textit{e.g.},][]{pad1993} of the global {dynamical properties of the galaxy} system. 

{The process of formation and evolution of a cluster from a matter {density} perturbation to a galaxy system in dynamical equilibrium is driven both by the initial cosmological conditions and by various physical and stochastic mechanisms \citep[\textit{e.g.},][]{bt2008}. Initially, both the homogeneous background and the matter perturbation expand with the Hubble flow; however, a fraction of this matter condenses into a set of galaxies that decouple from the expansion. The cluster, itself, {in the course of time,} decouples from the expansion, forming now a gravitationally bound system, which {turns around} and begins a process of collapse that ends in an eventual virialization \citep[\textit{e.g.},][]{gu1972}. This is a state of statistical equilibrium of internal {gravitational} forces, reflected in the averages of the total kinetic and potential energies of {galaxies} \citep[\textit{e.g.},][]{lib1960}.} 
Analyzing the {transitory} evolutionary state of {galaxy systems} is not a trivial task, {even when} gravity is the {sole driving force shaping} their evolutionary processes. Here, we use the {term} `evolutionary state' to refer to the degree of progress a galaxy system has in its evolutionary line, starting from its formation and ending at equilibrium (relaxation). 

In this work we {present a method to characterize} the evolutionary state of galaxy systems {by estimating the} entropy component that depends only on the macroscopic state of their galaxy ensembles. {For this, we propose a specific entropy estimator ($H_Z$) that combines (optical) observational parameters of the systems such as virial mass, volume, and galaxy velocity dispersion.} Our fundamental premise is that {a galaxy ensemble} should evolve in the sense of increasing entropy, modifying its distribution {of galaxies} in the observed phase-space {(which includes radial, angular and velocity coordinates)} to a more random and dynamically relaxed one where there are no macroscopic movements or special configurations \citep[\textit{e.g.},][]{ll1980,sas1984}. {This means that the spatial and velocity distributions of member galaxies change as the system evolves, starting from more substructured ensembles (with less entropy) towards more homogeneous ones in dynamical relaxation (with higher entropy). Of course, the parameters associated with the global dynamics of the system also change in the process, and may be useful for the estimation of state functions such as entropy.} No assumption is made here about the distribution of DM and ICM within the cluster \citep[\textit{e.g.},][]{lk2003,lim2005}, but only their important contribution to the total gravitational potential that determines the dynamics of the galaxy ensemble.

{To evaluate the $H_Z$-entropy estimator we make use of different tests. The first is the comparison between $H_Z$-entropy estimator and a discrete classification of assembling states applied to an observational samples of 70 nearby clusters ($z \lesssim 0.15$).} 
The second test is the calculation of the Shannon entropy, which provides a quantitative measure of the degree of disorder (or uncertainty) in the distribution of galaxies at different regions within each cluster phase-space. Substructures are considered as special configurations and their presence reduces the Shannon entropy, so more relaxed systems (with a more random galaxy distribution in the phase-space) are expected to have higher entropies. In this test, galaxy systems are considered as 
%---optical--- 
data sets whose information entropies can be calculated.
{Both the observational cluster sample and another from simulated cluster halos, with their respective subhalos, are used to compare $H_Z$ and Shannon entropies.}
{As complementary validations, different approaches are used, including the analysis of the relaxation probability of the systems (based on statistical distances between empirical and reference galaxy distributions) and the study of correlations between $H_Z$ and other 
%---discrete and 
continuous parameters commonly associated to the dynamical state of the systems.} 

In Section \ref{sec:equil}, we extend the discussion about the dynamical equilibrium and stability of galaxy clusters, {together with a brief description of the way evolutionary state is commonly characterized from observed and simulated data.} In Section \ref{sec:entropy}, we focus on the entropy-based estimator and present our proposal to quantify the dynamical state of galaxy systems.  In Section \ref{sec:test}, we apply our method to a sample of 70 well-sampled galaxy clusters, from very rich to poor ones, in the nearby Universe. In Section \ref{sec:shannon} we calculate the Shannon entropy for both the observational sample and a sample of 248 cluster halos from the IllustrisTNG simulation. Then, we compare this parameter with $H_Z$. Other dynamical parameters are evaluated in Section \ref{sec:valid} also for validating $H_Z$. Discussion and conclusions are presented in Section \ref{sec:disc}. 
Throughout this paper we assume a flat $\Lambda$CDM cosmology with the following parameters: Hubble constant $H_0=70$ km s$^{-1}$ Mpc$^{-1}$, matter density $\Omega_M=0.3$ and dark energy density $\Omega_{\Lambda}=0.7$. \\

%%%%%%%%%%%%%%%%%%%%%%%%%%%%%%%%%%%%%%%%%%%%%%%%%%%%%%%%%
% Section 2
%%%%%%%%%%%%%%%%%%%%%%%%%%%%%%%%%%%%%%%%%%%%%%%%%%%%%%%%%
\section{{Evolution,} equilibrium and stability of galaxy clusters} \label{sec:equil}

% Subection 2.1
%%%%%%%%%%%%%%%%%%%%%%%%%%%%%%%%%%%%%%%%%%%%%%%%%%%%%%%%%
\subsection{{Reaching} virial equilibrium}

{The evolution of galaxy systems at various scales is understood today through the} hierarchical formation model \citep[\textit{e.g.},][]{pe1980,pad1993,za2018} in a $\Lambda$CDM scenario.  {According to this model, the smallest systems (galaxy groups and poor clusters) merge ---via gravity--- to form the largest ones (rich clusters and superclusters), being the observed substructuring a consequence of this process. Subsequently, these galaxy ensembles undergo a non-linear gravitational collapse during which a combination of} physical and stochastic mechanisms \citep[\textit{e.g.},][]{ly1967,sas1980,pad1990} {drive the systems into a further state of dynamical relaxation ---the virial equilibrium--- in which their substructures vanish \citep[\textit{e.g.},][]{am2009}.} 

{Galaxy clustering is an irreversible process where, as galaxies accumulate, different mechanisms tend to increase the number of ways in which, in a statistical sense, internal energy can be distributed within the systems \citep[\textit{e.g.},][]{sas1980,sas1984}. Once bound and during collapse, the galaxy system undergoes internal ---dissipationless--- 
%and irreversible 
processes \citep[\textit{e.g.}, violent relaxation, phase mixing, 
%mixing of different galaxy species
energy equipartition, {galaxy mergers, fading of substructures and density or temperature gradients,} among others, 
%morphological segregation of galaxies, and even the gravitational clustering process itself \citep[\textit{e.g.},][]{sas1980,iq2011}.
see,][]{wt1996,deh2005,bt2008} that lead to states of greater dynamical relaxation.} 
All these processes are dominated by gravitational interactions that, during the relaxation time, tend to homogenize the spatial distribution of the member galaxies and distribute their radial velocities in a Gaussian way ---inside the clusters, galaxies are scattered randomly \citep[achieving a quasi-Maxwellian velocity distributions as {they tend to virialization}, \textit{e.g.},][]{sas1984,sam2014}, making the tendency for macroscopic motions to disappear. This requires the individualization of the motions of the galaxies, so that any substructure (\textit{e.g.}, accreted groups) will be `dissolved' before the cluster virializes. 
%, probably during collapse and through violent relaxation \citep[\textit{e.g.},][]{ly1967,wt1996}. }
The thermodynamic perturbations and the changes in the distribution of the ICM inside the clusters during their evolution also produce increases in the total entropy of these systems \citep[\textit{e.g.},][]{tn2001,voi2005}.

{The} equilibrium state can be understood, in first approximation, as that in which the gravitational collapse is supported by the effect of inertial ---centrifugal or dispersion--- forces, achieving relaxed internal configurations (called states of dynamical relaxation), {that is,} in which there are no unbalanced potentials, such as gravitational-driving forces, within galaxy systems. In this sense, collisionless systems in equilibrium are analogous to self-gravitating fluids because they support gravitational collapse through ``pressure gradients'' proportional to the velocity dispersion that, at each point, tend to disperse any local increase in particle density \citep[\textit{e.g.},][]{bt2008}. 
{Furthermore, dynamical equilibrium is characterized by the statistical equality of the cluster mass profiles obtained from different galaxy populations within the cluster \citep[\textit{e.g.},][]{car1997a}, implying that all these populations are in equilibrium with the cluster potential according to the hydrodynamical equilibrium model by Jeans.} 

Throughout the evolution of an isolated cluster, its total internal energy $U=K+W$ is conserved so that, as the member galaxies get closer together (spontaneous reduction of inter-particle distance $r_{ij}$ by mutual attraction), the gravitational potential energy $W$ decreases (becoming more negative) and consequently the total internal kinetic energy $K$ increases. This is reflected observationally as an increase in the velocity dispersion of galaxies, which can stop the collapse. Thus, when the system finds some route to dynamical equilibrium, the increase in $K$ at the expense of the decrease in $W$ does not continue indefinitely, but evolves toward configurations in which the ratio
\begin{equation}\label{b}
b=-\frac{W}{2K},
\end{equation}

\noindent tends to one \citep[$b\rightarrow 1$, \textit{e.g.},][]{sas1984,bt2008}. The asymptotic limit, $b=1$, of this tendency is the virial equilibrium: 
\begin{equation}\label{VT}
W=-2K,
\end{equation}

\noindent a state in which dynamical parameters that characterize the global configuration of the system remain, at least temporarily, stationary (relaxed). The virial theorem expresses a statistical equilibrium between the temporal averages of the total internal kinetic and gravitational potential energies, \textit{i.e.}, $\langle W\rangle_\tau=-2\langle K\rangle_\tau$, but these energies are also the result of an ensemble of interacting galaxies \citep[\textit{e.g.},][]{lib1960}, so it is reasonable to approximate the temporal averages by the average of the ensemble expressed in (\ref{VT}). Virial equilibrium is assumed to be the type of dynamical equilibrium reached by self-gravitating systems.

% Subsection 2.2
%%%%%%%%%%%%%%%%%%%%%%%%%%%%%%%%%%%%%%%%%%%%%%%%%%%%%%%%%
\subsection{Stability}

Going beyond the description of equilibrium in galaxy systems, we need to discuss if this equilibrium is stable and this topic requires involving the concept of `entropy'.
As any thermodynamic system, self-gravitating systems progress in the sense of increasing entropy \citep[\textit{e.g.},][]{lp1981,thl1986,pg2013} towards the state of dynamical equilibrium described above. {Galaxy clustering simulations also confirm this fact \citep[\textit{e.g.},][]{sas1984,iq2006,iq2011}.}  
{Due to the scattering experienced by galaxies in the phase-space of clusters}, {these become the regions within the LSS where first-order entropy production occurs by increasing the randomness of the motion of galaxies during their gravitational accumulation. Even if galaxy clusters are isolated, some entropy is generated ---or produced--- due to the presence of internal irreversibilities.}
The peculiarity here is that the virial equilibrium is not unique, but only a \emph{metastable equilibrium state} \citep[\textit{e.g.},][]{an1962,ly1968,pad1990,cha2002}. This means that the entropy of self-gravitating systems can grow indefinitely without reaching a global maximum, that is, the virial equilibrium is only a state of local extreme of entropy \citep[\textit{e.g.},][]{pad1989}.

The dynamical equilibrium of a galaxy cluster can be disturbed if it actively interacts with its surroundings, for example through mergers with other clusters and/or group accretions {or tidal forces}. As a result, the cluster {takes a route towards a new equilibrium, in a state} of higher entropy. 
{Concerning the impact of the interaction}, {if the accreted groups are very small, the clusters can be kept unperturbed in states close to equilibrium. In more extreme cases, the merger of two massive clusters completely removes the systems from their equilibrium.
%, directing their evolution towards a new equilibrium state of higher entropy. 
In dense environments, such as supercluster cores, the accretion of galaxies and groups by the most massive clusters continually disturbs their dynamical states. On the other hand, in less dense environments, such as {along} filaments or edges near voids, clumpy clusters evolve as quasi-isolated systems, reaching dynamical relaxation {possibly} faster, without many significant disturbances, {but accessing lower entropy levels compared to clusters in ``busy'' environments}. 
{That is,} depending on the cosmological environment inside the LSS in which a cluster evolves, its relaxation process may be affected several times ---or not--- given the amount of matter available in its surroundings.}  

The most stable states of a galaxy cluster ---of mass $\mathcal{M}$ and radius $R$--- are favored when the density contrast between its center and its edge is $\rho_0/\rho(R)<709$ and the internal energy is $U>-0.335G\mathcal{M}^2/R$ \citep[the Antonov instability or gravothermal catastrophe, \textit{e.g.},][]{an1962,ly1968}, and under these conditions the virial equilibrium corresponds to a local maximum of entropy \citep[\textit{e.g.},][]{pad1989,pad1990}. In fact, the most stable dynamical configurations that self-gravitating systems can access are those in which the particle {---or matter---} distribution settles on a core-halo structure \citep[\textit{e.g.},][]{bt2008,cha2002}. This {theoretical} structure, characterized by a collapsed core coexisting with a regular halo, has also been confirmed by simulations \citep[\textit{e.g.},][]{co1980,bsi2002} {and recognized by the distributions of galaxies \citep[\textit{e.g.},][]{sa1988,ad1998} and the ICM in observed clusters \citep[\textit{e.g.},][]{tn2001,cav2009}}. \\

% Subection 2.3
%%%%%%%%%%%%%%%%%%%%%%%%%%%%%%%%%%%%%%%%%%%%%%%%%%%%%%%%%
\subsection{{Estimating the evolutionary state of galaxy systems}}

{Observationally, a cluster close to dynamical equilibrium is distinguished from a non-relaxed one by exhibiting a more regular morphology \citep[\textit{e.g.},][]{sa1988}, {both in the optical and X-rays,} and a more homogeneous ---projected--- spatial distribution of member galaxies \citep[without the presence of significant substructures, \textit{e.g.},][and references therein]{ca2021}, as well as by having a more isotropic galaxy velocity distribution \citep[or Gaussian in the line-of-sight, \textit{e.g.},][]{gi2001,sam2014}.} 

{Concerning global morphology,} {the most dynamically relaxed clusters tend to present} {low ellipticity shapes in the projected distribution of galaxies and X-ray 
%emission 
surface brightness maps, with a ---possible--- single peak at their centers.}
%\citep[\textit{e.g.},][]{jo1984, sa1988,Lag19,ca2021}.
%The observed distribution of member galaxies or of the ICM within the clusters reflects the proximity of these to the equilibrium states. 
{Several morphological classifications have been proposed following this premise \citep[\textit{e.g.},][and references therein]{sa1988}.}
{There are also indicators of the internal structure of the galaxy systems, such as 
%In practice, our understanding about the internal dynamics of galaxy clusters relies mainly on the analysis of observational parameters 
the radial profile and the degree of concentration of galaxies \citep[\textit{e.g.},][]{ad1998,tl2015,ka2020}, 
%the shape and {velocity dispersion} \citep[\textit{e.g.},][]{sam2014,ca2021}, 
as well as {a measure of} the presence of substructures within {them} using 1D, 2D and 3D tests \citep[\textit{e.g.},][]{GeB1982,dr1988,ca2021}.}

{In this sense,} {a significantly substructured system, either {from} optical observation of galaxy subclumps \citep[\textit{e.g.},][]{GeB1982,br2009,ca2021} or the detection of multiple peaks in X-ray emission \citep[\textit{e.g.},][]{jo1984,bu1995,Lag19}, cannot be considered in {dynamical} equilibrium. 
%show the distribution of hot gas within the clusters and may reveal the presence of secondary peaks of X-ray emission possibly related to ICM trapped in substructures. 
Thus, the presence {and significance} of substructures %{---the assembly state of the galaxy system---} 
reveal {the far the galaxy system is from a relaxed and homogeneous global potential. A }
description of the cluster level of internal substructuring {is called} 
%characteristics such as its morphology or 
its gravitational assembly state \citep[\textit{e.g.},][]{ca2021}.}

{In principle,} {one can also measure global}
%the evolutionary state can be studied by estimating 
{parameters related to the internal dynamics \citep[\textit{e.g.},][]{car1996,gi2001} ---or dynamical state--- of the galaxy system.}
{These parameters are associated, for example, with the mass, radius and velocity dispersion of the system.}
%, as well as} {observe} 
%by observing 
{Such approach} {supposes that both the observable aspect {and structure} and the dynamical parameters of the system change in a correlated manner during its evolution, dominated by mechanisms that take it from more irregular and substructured configurations to those with more homogeneous galaxy distributions and more dynamically relaxed \citep[\textit{e.g.},][]{am2009}.} 

{Furthermore, from X-ray {observations one can} construct entropy (or temperature or density) profiles that account for the {evolutionary history, structure} and thermodynamic state of the ICM inside the clusters \citep[\textit{e.g.},][]{tn2001,voi2005,cav2009}, which {also} contributes to the study of the {degree} of relaxation ---or disturbance--- of their gravitational potentials} {(analysis of self-similarity of clusters)}. {Nevertheless, we are not always fortunate enough to detect X-ray emission from clusters {nor to have the amount of data necessary to carry out massive studies}, so for this we are still limited to optical surveys.} \\

%%%%%%%%%%%%%%%%%%%%%%%%%%%%%%%%%%%%%%%%%%%%%%%%%%%%%%%%%
% Section 3
%%%%%%%%%%%%%%%%%%%%%%%%%%%%%%%%%%%%%%%%%%%%%%%%%%%%%%%%%
%\section{Assembly state of galaxy clusters} \label{sec:assembly}

\section{Entropy {of galaxy systems from global parameters}} \label{sec:entropy}

% Subsection 3.1
%%%%%%%%%%%%%%%%%%%%%%%%%%%%%%%%%%%%%%%%%%%%%%%%%%%%%%%%%
%\subsection{An entropy estimator for galaxy ensembles} \label{sec:s}

Based on the `classical' concept of entropy of a particle system, it is possible to construct an estimator for the entropy component related to the set of member galaxies of a cluster, the `galaxy ensemble'.
The depth of the cluster's global potential well, which determines how fast the bound galaxies must move, is proportional to the total mass of the cluster that can be estimated, with negligible bias \citep[see,][]{bv2006}, by the virial mass estimator
\begin{equation}\label{M_vir} 
\mathcal{M}_{\mathrm{vir}}=\frac{\alpha \pi}{2G}\sigma_{LOS}^2 R_{\mathrm{p}},
\end{equation}
where $\sigma_{LOS}$ is the line-of-sight {(LOS)} velocity dispersion of the sampled galaxies, $\alpha$ is a deprojection parameter for $\sigma_{LOS}$ assuming anisotropy in the galaxy velocity distribution \citep[$\alpha=3$ if orbits have an isotropic and isothermal distribution, \textit{e.g.},][and references therein]{tl2015}, and 
\begin{equation}\label{R_p}
R_{\mathrm{p}}=\frac{N(N-1)}{\sum_{k<l} 1/R_{kl}},
\end{equation} %\left( \right) \frac{1}{R_{kl}
is the projected mean radius of the distribution of cluster galaxies, where $R_{kl}$ is the projected distance (in Mpc) between pairs of the $N$ sampled galaxies. The factor $\pi/2$ in (\ref{M_vir}) is the radius deprojection factor \citep[\textit{e.g.},][]{lib1960} so that statistically $R_{\mathrm{vir}}\simeq \pi R_{\mathrm{p}}/2$ is the three-dimensional virial radius of the cluster \citep[\textit{e.g.}, ][]{car1996,gi2001}, a measure of the region of gravitational confinement of its member galaxies. 

Now, in a first approximation, we can imagine the set of cluster galaxies as a system of particles with mean kinetic energy \citep[\textit{e.g.},][]{sc2015} 
\begin{equation}\label{K}
K=\frac{\alpha}{2}\mathcal{M}_{\mathrm{vir}}\sigma_{LOS}^2,  
\end{equation}
and confined in a region of volume 
\begin{equation}\label{V}
V=\frac{4}{3}\pi R_{\mathrm{vir}}^3.   
\end{equation}

We assume that the velocity dispersion of galaxies is proportional to the `temperature' $T$ of the galaxy ensemble\footnote{Thought, simplistically, as a system of point masses (particles).}, so that $\sigma_{LOS}^2=\beta T$, with $\beta$ being a proportionality constant that transforms temperature units into square velocity units. Furthermore, the mean `pressure' of the galaxy ensemble, commonly approximated as $P=\bar{\rho}\,\sigma_{LOS}^2$, where $\bar{\rho}$ is the mean mass density of the system \citep[\textit{e.g.},][]{sc2015}, can be conveniently modified {to} the form $P=\bar{\rho}\,\sigma_{LOS}^2(1-2b)$, where the factor $(1-2b)$ has been introduced to generalize the expression \citep[see, for example,][]{sas1984}. Thus, $P>0$ for unbound systems ($b=0$), just like an ordinary gas of particles enclosed in a hypothetical sphere of volume $V$ with no gravitational potential; while $P<0$ in self-gravitating bound systems ($b>0.5$). The mean value of pressure in a marginal virial equilibrium ($b\longrightarrow 1$), for a system with $\bar{\rho}=\mathcal{M}_{\mathrm{vir}}/V$, is 
\begin{equation}\label{P}
P=-\bar{\rho}\,\sigma_{LOS}^2=-\frac{2K}{\alpha V},
\end{equation}
matching the definition of the so-called gravitational pressure \citep[\textit{e.g.},][]{pad2000}, $P=W/3V$, when $\alpha\longrightarrow 3$.

In virial equilibrium, the internal energy of the galaxy ensemble is $U=K+W=-K$, according to (\ref{VT}). However, for any state of the system including those prior to equilibrium, the internal energy can be generalized, as in \citet{sas1984}, in the form $U=K(1-2b)$, where, for unbound systems ($b=0$) the internal energy is only kinetic, while for bound and virialized systems ($b\longrightarrow 1$) we get the marginal value
\begin{equation}\label{U}
U=-\frac{\alpha}{2} \mathcal{M}_{\mathrm{vir}}\beta T,
\end{equation}
where $\sigma_{LOS}^2=\beta T$, as before. Note that the mean potential energy of the galaxy system does not appear explicitly in its ``thermodynamic description''. 

As can be inferred from (\ref{M_vir}), galaxy systems of fixed mass internally heat up ($T$ increases) when they contract ($R_{\mathrm{vir}}$ decreases) and cool down ($T$ decreases) when they expand ($R_{\mathrm{vir}}$ increases). In addition, from (\ref{U}) it is possible to appreciate an atypical behavior of virialized self-gravitating systems. If we allow the galaxy systems to exchange energy ---but not matter--- with the environment, then they cool down ($dT<0$) by receiving energy ($dU>0$) from the environment and warm up ($dT>0$) by releasing energy ($dU<0$) to it. These types of systems are said to have negative specific heats \citep[$dU/dT<0$, \textit{e.g.},][]{ly1977}. 

A fundamental expression of the form $u=u(s,\upsilon)$, which relates the specific variables $u$, $s$ and $\upsilon$ of internal energy, entropy and volume respectively, must be satisfied by a single-component thermodynamic system \citep[\textit{e.g.},][]{sas1984}. Differentiating, we get that $du=(\partial u/\partial s)_{\upsilon}ds+(\partial u /\partial \upsilon)_s d\upsilon$, obtaining the standard Gibbs $Tds$ relation $du=Tds-Pd\upsilon$, with $T\equiv (\partial u/\partial s)_{\upsilon}$ and $P\equiv -(\partial u /\partial \upsilon)_s$ being {respectively} the temperature and pressure of the system. 
Let $u=U/\mathcal{M}_{\mathrm{vir}}$, $\kappa=K/\mathcal{M}_{\mathrm{vir}}$, $s=S/\mathcal{M}_{\mathrm{vir}}$ and $\upsilon=V/\mathcal{M}_{\mathrm{vir}}$ the specific variables per unit of mass for internal and  kinetic energies, entropy and volume, respectively. It is evident that Gibbs $Tds$ relation
%, $du=Tds-Pd\upsilon$, 
does not fit in its original form to self-gravitating systems. For ordinary systems of particles, both adiabatic compression ($d\upsilon<0$) and isocoric heat input from the environment ($Tds>0$) imply an increase in their internal energy ($du>0$). Instead, for self-gravitating system of particles, compressions imply ``heating'' ($dT,d\kappa>0$), but with a decrease in internal energy ($du<0$), which leads to an increase in the entropy of the system ($ds>0$) according to the direction of the spontaneous process of gravitational accumulation.

It is necessary to use an expression analogous to the Gibbs $Tds$ equation, but which conforms to the thermodynamic behaviour of self-gravitating systems described above. In the considered galaxy ensemble the entropy increases along with the internal kinetic energy as they virialize (see, section \ref{sec:equil}). Then, we can impose that, in systems of point galaxies, $s=s(\kappa,\upsilon)$ with differential form 
\begin{equation}\label{dS}
ds=\left(\frac{\partial s}{\partial \kappa}\right)_\upsilon d\kappa+\left(\frac{\partial s}{\partial \upsilon}\right)_{\kappa} d\upsilon,
\end{equation}
where, by analogy with the well-known thermodynamic expressions for temperature and pressure in the Gibbs $Tds$ equation, we have
\begin{equation}\label{TyP}
\left(\frac{\partial s}{\partial \kappa}\right)_\upsilon=\frac{1}{T},\,\ \mathrm{and} \,\ \left(\frac{\partial s}{\partial \upsilon}\right)_\kappa=\frac{P}{T}.
\end{equation}   

Note that, for self-gravitating galaxy systems in general, we need the $1/T>0$ and $P/T<0$ conditions to be satisfied. The last condition is required to obtain entropy increases during the gravitational collapse processes (in which the systems undergo contractions, $d\upsilon<0$). This suggests a negative pressure in this type of systems \citep[like systems in phase transitions or metastable states in liquids, \textit{e.g.},][]{im2007} that causes a ``repulsion'' and prevents a singular collapse, just as proposed by the expression (\ref{P}). 

Finally, in order to obtain an estimator for the entropy of galaxy clusters, we will solve the system of differential equations (\ref{TyP}) assuming that $1/T=\alpha \beta/2\kappa$ and $P/T=-\beta/\upsilon$, obtained by combining (\ref{K}) and (\ref{P}). Thus, the solution can be verified to be of the form
\begin{equation}\label{s}
s(\kappa,\upsilon)= \beta\ln{\left(\kappa^{\frac{\alpha}{2}}\upsilon^{-1}\right)}+s_0, 
\end{equation}
where $s_0$ is an integration constant possibly related to the initial entropy of the galaxy ensemble, \textit{e.g.}, the entropy it had when the distribution of galaxies was not yet concentrated before gravitational clustering (see, section \ref{sec:equil}). Replacing the expressions $\kappa=(\alpha/2)\sigma_{LOS}^2$ and $\upsilon=(4/3)\pi R_{\mathrm{vir}}^3\mathcal{M}_{\mathrm{vir}}^{-1}$  in (\ref{s}) we get the dimensionless estimator
\begin{equation}\label{s2}
H_Z\equiv \frac{s-s_0}{\beta}=
\ln{\left( \frac{\mathcal{M}_{\mathrm{vir}}}{\frac{4}{3}\pi R_{\mathrm{vir}}^3} \right)} +
 \frac{\alpha}{2}\ln{\left( \frac{\alpha}{2} \sigma_{LOS}^2 \right)},
\end{equation}
defined only in terms of observational parameters {that can be obtained through optical data (\textit{e.g.}, galaxy coordinates and redshifts)}.\\

%%%%%%%%%%%%%%%%%%%%%%%%%%%%%%%%%%%%%%%%%%%%%%%%%%%%%
% Section 4
%%%%%%%%%%%%%%%%%%%%%%%%%%%%%%%%%%%%%%%%%%%%%%%%%%%%%%%%%
\section{Testing the $H_Z$ entropy estimator in galaxy systems} \label{sec:test}

% Subsection 4.1
%%%%%%%%%%%%%%%%%%%%%%%%%%%%%%%%%%%%%%%%%%%%%%%%%%%%%%%%%
\subsection{Observational data} \label{ssec:observ}

We use data from \citet{ca2021}, a sample of 67 galaxy clusters, %mostly 
from Abell/ACO \citep{ab1958,ab1989} {catalogs}, with redshifts up to $z \sim 0.15$. These clusters were selected because they are among the most well sampled galaxy systems in the nearby Universe, and cover roughly uniformly from poor to rich systems (with ICM-temperatures from 1 to 12 keV), being balanced for including all BM \citep{bm1970} types. Three other non-Abell clusters with {similar} characteristics were included in our sample {(AM0227-334, SC1329-313 and MKW03S)}, which allow us to call, for short, our observational sample \emph{Top70} from now on {(Table \ref{tab:init})}. For each cluster, a sample of spectroscopically confirmed member galaxies is available, selected inside the cluster caustics up to {a fiducial} aperture of $1.3\times r_{200}$ from the centre of the cluster (chosen to be the 
%First Ranked Galaxy, FRG).
{Central Dominant Galaxy, CDG}).
%, taking $r_{200}$ as in \citet{car1997b}:
%\begin{equation}
%    r_{200}=\frac{\sqrt{3}\sigma_{LOS}}{10H(z)}
%\end{equation}
%where $H(z)=H_0\left[\Omega_r(1+z)^4 +\Omega_m(1+z)^3 +\Omega_k(1+z)^2 +\Omega_\Lambda \right]^{1/2}$ is the Hubble-Lema\^i{}tre parameter at redshift $z$, and taking $\Omega_r\sim 0$ and $\Omega_k \sim 0$. 
{The numbers of these presumably virialized members ranges from 21 to 919 (average 154), while originally at least 90 spectroscopic redshifts were available for each cluster \citep[see][for the details of the process for determining membership, caustics and virial radius]{ca2021}. One should note that the richness of a cluster is related to both intrinsic and observational conditions ---more massive clusters are richer, while nearby clusters tend to be preferentially observed.} {However, the numbers of members we have are} adequate for minimizing observational biases in counting them in bins. Astrometric positions of galaxies have uncertainties of $\pm 0.25''$, and radial velocities of $\pm 60$ km s$^{-1}$ \citep[see,][and references therein]{ca2021}.

%TABLAS
\begin{table*}
\centering
\caption{Cluster sample (Top70).} \label{tab:init} 
\resizebox{12cm}{!}{
%\hspace*{-3.5cm}
\begin{tabular}{lrrrr c rr c rrr}
\toprule %\hline \hline
%---------------------------------------------------------------------------------
   \multicolumn{5}{c}{Optical data} & & 
   \multicolumn{2}{c}{X-ray data$^\mathrm{b}$} & &
   \multicolumn{3}{c}{Basic properties} \\
   \cline{1-5} \cline{7-8} \cline{10-12}
  \multicolumn{1}{c}{Name$^\mathrm{a}$} &
  \multicolumn{1}{c}{RA$_\mathrm{CDG}$} &
  \multicolumn{1}{c}{Dec$_\mathrm{CDG}$} &
  \multicolumn{1}{c}{$\bar{z}$} & 
  \multicolumn{1}{c}{$N_a$} & &
  \multicolumn{1}{c}{$r_{500}$} &
  \multicolumn{1}{c}{$k\mathrm{T}_X$} & &
  \multicolumn{1}{c}{$\sigma_{LOS}$} &
  \multicolumn{1}{c}{$\mathcal{M}_\mathrm{vir}$} &
  \multicolumn{1}{c}{$R_\mathrm{vir}$} \\

  \multicolumn{1}{c}{} &
  \multicolumn{1}{c}{[deg]$_\mathrm{J2000}$} &
  \multicolumn{1}{c}{[deg]$_\mathrm{J2000}$} &
  \multicolumn{1}{c}{} & 
  \multicolumn{1}{c}{} & &
  \multicolumn{1}{c}{[Mpc]} &
  \multicolumn{1}{c}{[keV]} & &
  \multicolumn{1}{c}{[km/s]} &
  \multicolumn{1}{c}{[$10^{14}\mathcal{M}
  _\odot$]} &
  \multicolumn{1}{c}{[Mpc]} \\
    
  \multicolumn{1}{c}{(1)} &
  \multicolumn{1}{c}{(2)} &
  \multicolumn{1}{c}{(3)} &
  \multicolumn{1}{c}{(4)} & 
  \multicolumn{1}{c}{(5)} & &
  \multicolumn{1}{c}{(6)} &
  \multicolumn{1}{c}{(7)} & &
  \multicolumn{1}{c}{(8)} & 
  \multicolumn{1}{c}{(9)} &
  \multicolumn{1}{c}{(10)}\\
\hline
    A2798B &   9.37734 & -28.52947 & 0.1119 &  60 & & 0.7476 &  3.39 & &  757  &  6.01 & 1.75 \\
    A2801  &   9.62877 & -29.08160 & 0.1122 &  35 & &    ... &  3.20 & &  699  &  6.94 & 1.83 \\
    A2804  &   9.90754 & -28.90620 & 0.1123 &  48 & &    ... &  1.00 & &  516  &  3.11 & 1.40 \\
   A0085A  &  10.46052 &  -9.30304 & 0.0553 & 318 & & 1.2103 &  7.23 & & 1034  & 19.75 & 2.65 \\
    A2811B &  10.53718 & -28.53577 & 0.1078 & 103 & & 1.0355 &  5.89 & &  947  & 13.94 & 2.32 \\
    A0118  &  13.75309 & -26.36238 & 0.1144 &  72 & &    ... &   ... & &  680  &  6.16 & 1.76 \\
    A0119  &  14.06709 &  -1.25549 & 0.0444 & 294 & & 0.9413 &  5.82 & &  853  &  9.51 & 2.08 \\
    A0122  &  14.34534 & -26.28134 & 0.1136 &  28 & & 0.8165 &  3.70 & &  677  &  4.98 & 1.64 \\
    A0133A &  15.67405 & -21.88215 & 0.0562 &  86 & & 0.9379 &  4.25 & &  778  &  7.30 & 1.90 \\
    A2877-70  &  17.48166 & -45.93122 & 0.0238 & 112 & & 0.6249 &  3.28 & &  679  &  4.20 & 1.60 \\
AM0227-334 & 37.33891  & -33.53196 & 0.0780 &  30 & &    ... &   ... & &  625  &  4.11 & 1.56 \\
    A3027A &  37.70601 & -33.10375 & 0.0784 &  82 & & 0.7200 &  3.12 & &  713  &  7.52 & 1.90 \\
    A0400  &  44.42316 &   6.02700 & 0.0232 &  51 & & 0.6505 &  2.25 & &  343  &  0.68 & 0.87 \\
    A0399  &  44.47120 &  13.03080 & 0.0705 &  69 & & 1.1169 &  6.69 & &  950  & 11.65 & 2.21 \\
    A0401  &  44.74091 &  13.58287 & 0.0736 & 114 & & 1.2421 &  7.06 & & 1026  & 15.11 & 2.41 \\
    A3094A &  47.85423 & -26.93122 & 0.0685 &  84 & & 0.6907 &  3.15 & &  637  &  4.83 & 1.65 \\
    A3095  &  48.11077 & -27.14017 & 0.0652 &  21 & &    ... &   ... & &  327  &  0.65 & 0.84 \\
    A3104  &  48.59055 & -45.42024 & 0.0723 &  28 & & 0.8662 &  3.56 & &  498  &  1.77 & 1.18 \\
    S0334  &  49.08556 & -45.12110 & 0.0746 &  26 & &    ... &   ... & &  534  &  2.11 & 1.25 \\
    S0336  &  49.45997 & -44.80069 & 0.0773 &  32 & &    ... &   ... & &  538  &  3.02 & 1.40 \\
    A3112B &  49.49025 & -44.23821 & 0.0756 &  74 & & 1.1288 &  5.49 & &  705  &  8.45 & 1.98 \\
    A0426A &  49.95098 &  41.51168 & 0.0176 & 314 & & 1.2856 &  6.42 & & 1029  & 13.50 & 2.36 \\
    S0373  &  54.62118 & -35.45074 & 0.0049 &  98 & & 0.4017 &  1.56 & &  390  &  0.43 & 0.75 \\
    A3158  &  55.72063 & -53.63130 & 0.0592 & 249 & & 1.0667 &  5.42 & & 1066  & 13.93 & 2.35 \\
    A0496  &  68.40767 & -13.26196 & 0.0331 & 279 & & 0.9974 &  4.64 & &  712  &  6.31 & 1.82 \\
    A0539  &  79.15555 &   6.44092 & 0.0288 &  92 & & 0.7773 &  3.04 & &  698  &  3.92 & 1.56 \\
    A3391  &  96.58521 & -53.69330 & 0.0560 &  75 & & 0.8978 &  5.89 & &  817  &  7.74 & 1.94 \\
    A3395  &  96.90105 & -54.44936 & 0.0496 & 199 & & 0.9298 &  5.10 & &  746  &  6.42 & 1.82 \\
    A0576  & 110.37600 &  55.76158 & 0.0379 & 191 & & 0.8291 &  4.27 & &  866  & 11.18 & 2.20 \\
    A0634  & 123.93686 &  58.32109 & 0.0268 &  70 & &    ... &   ... & &  395  &  1.13 & 1.03 \\
    A0754  & 137.13495 &  -9.62974 & 0.0542 & 333 & & 1.1439 &  8.93 & &  820  &  9.10 & 2.05 \\
    A1060  & 159.17796 & -27.52858 & 0.0123 & 343 & & 0.7015 &  2.79 & &  678  &  3.99 & 1.57 \\
    A1367  & 176.00905 &  19.94982 & 0.0215 & 226 & & 0.9032 &  3.81 & &  597  &  3.76 & 1.54 \\
    A3526A & 192.20392 & -41.31167 & 0.0100 & 126 & & 0.8260 &  3.40 & &  564  &  2.43 & 1.34 \\
    A3526B & 192.51645 & -41.38207 & 0.0155 &  45 & &    ... &   ... & &  317  &  0.44 & 0.75 \\
    A3530  & 193.90001 & -30.34749 & 0.0536 &  94 & & 0.8043 &  3.62 & &  631  &  4.63 & 1.63 \\
    A1644  & 194.29825 & -17.40958 & 0.0470 & 288 & & 0.9944 &  5.25 & & 1008  & 13.98 & 2.36 \\
    A3532  & 194.34134 & -30.36348 & 0.0557 &  58 & & 0.9201 &  4.63 & &  443  &  1.66 & 1.16 \\
    A1650  & 194.67290 &  -1.76139 & 0.0842 & 146 & & 1.1015 &  5.72 & &  723  &  7.55 & 1.90 \\
    A1651  & 194.84383 &  -4.19612 & 0.0849 & 158 & & 1.1252 &  7.47 & &  876  & 12.48 & 2.25 \\
    A1656  & 194.89879 &  27.95939 & 0.0233 & 919 & & 1.1378 &  7.41 & &  995  & 15.66 & 2.47 \\
    A3556  & 201.02789 & -31.66996 & 0.0482 &  90 & &    ... &  3.08 & &  520  &  2.59 & 1.35 \\
    A1736A & 201.68378 & -27.43940 & 0.0350 &  36 & & 0.9694 &  3.34 & &  386  &  1.30 & 1.08 \\
    A1736B & 201.86685 & -27.32468 & 0.0456 & 126 & &    ... &   ... & &  844  &  8.82 & 2.03 \\
    A3558  & 201.98702 & -31.49547 & 0.0483 & 469 & & 1.1010 &  5.83 & &  955  & 15.75 & 2.46 \\
SC1329-313 & 202.86470 & -31.82058 & 0.0448 & 46  & &    ... &   ... & & 383   &  1.01 & 0.99 \\
    A3562  & 203.39475 & -31.67227 & 0.0486 &  82 & & 0.9265 &  5.10 & &  594  &  3.94 & 1.55 \\
    A1795  & 207.21880 &  26.59301 & 0.0630 & 154 & & 1.2236 &  6.42 & &  780  &  7.09 & 1.88 \\
    A2029  & 227.73377 &   5.74491 & 0.0769 & 155 & & 1.3344 &  8.45 & &  931  &  7.82 & 1.93 \\
    A2040B & 228.19782 &   7.43426 & 0.0451 & 104 & &    ... &  2.41 & &  627  &  4.77 & 1.65 \\
\hline
\end{tabular}}
\end{table*}

%%%%%%%%%%%%%%%%
\begin{table*}[!ht]
\centering
{TABLE \ref{tab:init} ---\textit{continued}} 
\resizebox{12cm}{!}{
\begin{tabular}{lrrrr c rr c rrr}
\toprule %\hline \hline
%---------------------------------------------------------------------------------
   \multicolumn{5}{c}{Optical data} & & 
   \multicolumn{2}{c}{X-ray data$^\mathrm{b}$} & &
   \multicolumn{3}{c}{Basic properties} \\
   \cline{1-5} \cline{7-8} \cline{10-12}
  \multicolumn{1}{c}{Name$^\mathrm{a}$} &
  \multicolumn{1}{c}{RA$_\mathrm{CDG}$} &
  \multicolumn{1}{c}{Dec$_\mathrm{CDG}$} &
  \multicolumn{1}{c}{$\bar{z}$} & 
  \multicolumn{1}{c}{$N_a$} & &
  \multicolumn{1}{c}{$r_{500}$} &
  \multicolumn{1}{c}{$k\mathrm{T}_X$} & &
  \multicolumn{1}{c}{$\sigma_{LOS}$} &
  \multicolumn{1}{c}{$\mathcal{M}_\mathrm{vir}$} &
  \multicolumn{1}{c}{$R_\mathrm{vir}$} \\

  \multicolumn{1}{c}{} &
  \multicolumn{1}{c}{[deg]$_\mathrm{J2000}$} &
  \multicolumn{1}{c}{[deg]$_\mathrm{J2000}$} &
  \multicolumn{1}{c}{} & 
  \multicolumn{1}{c}{} & &
  \multicolumn{1}{c}{[Mpc]} &
  \multicolumn{1}{c}{[keV]} & &
  \multicolumn{1}{c}{[km/s]} &
  \multicolumn{1}{c}{[$10^{14}\mathcal{M}_\odot$]} &
  \multicolumn{1}{c}{[Mpc]} \\
    
  \multicolumn{1}{c}{(1)} &
  \multicolumn{1}{c}{(2)} &
  \multicolumn{1}{c}{(3)} &
  \multicolumn{1}{c}{(4)} & 
  \multicolumn{1}{c}{(5)} & &
  \multicolumn{1}{c}{(6)} &
  \multicolumn{1}{c}{(7)} & &
  \multicolumn{1}{c}{(8)} & 
  \multicolumn{1}{c}{(9)} &
  \multicolumn{1}{c}{(10)}\\
\hline
    A2052  & 229.18536 &   7.02167 & 0.0347 & 120 & & 0.9465 &  2.88 & &  648  &  4.28 & 1.60 \\
    MKW03S & 230.46613 &   7.70888 & 0.0443 &  75 & &    ... &   ... & &  607  &  3.46 & 1.49 \\
    A2065  & 230.62053 &  27.71228 & 0.0730 & 168 & & 1.0480 &  6.59 & & 1043  & 17.01 & 2.50 \\
    A2063A & 230.77210 &   8.60918 & 0.0345 & 142 & & 0.9020 &  3.34 & &  762  &  6.18 & 1.81 \\
    A2142  & 239.58345 &  27.23335 & 0.0902 & 157 & & 1.3803 & 11.63 & &  828  & 11.06 & 2.16 \\
    A2147  & 240.57086 &  15.97451 & 0.0363 & 397 & & 0.9351 &  4.26 & &  935  & 15.70 & 2.47 \\
    A2151  & 241.28754 &  17.72997 & 0.0364 & 276 & & 0.7652 &  2.10 & &  768  &  8.35 & 2.00 \\
    A2152  & 241.37175 &  16.43579 & 0.0443 &  64 & & 0.5783 &  2.41 & &  406  &  1.56 & 1.14 \\
    A2197  & 246.92114 &  40.92690 & 0.0304 & 185 & & 0.5093 &  2.21 & &  573  &  4.21 & 1.59 \\
    A2199  & 247.15949 &  39.55138 & 0.0303 & 459 & & 1.0040 &  4.04 & &  779  &  7.21 & 1.91 \\
    A2204A & 248.19540 &   5.57583 & 0.1518 &  38 & & 1.3998 & 10.24 & & 1101  & 20.33 & 2.59 \\
    A2244  & 255.67697 &  34.06010 & 0.0993 & 102 & & 1.1295 &  5.99 & & 1161  & 18.35 & 2.55 \\
    A2256  & 256.11353 &  78.64056 & 0.0586 & 280 & & 1.1224 &  8.23 & & 1222  & 20.63 & 2.68 \\
    A2255  & 258.11981 &  64.06070 & 0.0805 & 179 & & 1.0678 &  7.01 & & 1000  & 16.44 & 2.47 \\
    A3716  & 312.98715 & -52.62983 & 0.0451 & 123 & &    ... &  2.19 & &  783  &  6.99 & 1.88 \\
    S0906  & 313.18958 & -52.16440 & 0.0482 &  26 & &    ... &   ... & &  440  &  1.46 & 1.11 \\
    A4012A & 352.96231 & -34.05553 & 0.0542 &  39 & &    ... &   ... & &  575  &  3.15 & 1.44 \\
    A2634  & 354.62244 &  27.03130 & 0.0309 & 166 & & 0.7458 &  3.71 & &  717  &  5.87 & 1.78 \\
    A4038A-49 & 356.93768 & -28.14071 & 0.0296 & 180 & & 0.8863 &  2.84 & &  753  &  5.95 & 1.79 \\
    A2670  & 358.55713 & -10.41900 & 0.0760 & 251 & & 0.9113 &  4.45 & &  970  &  9.91 & 2.09 \\
\hline
\end{tabular}}
\begin{flushleft}  
\footnotesize{$^\mathrm{a}$ A capital letter after the ACO name indicates the line-of-sight component of the cluster.}\\
\footnotesize{$^\mathrm{b}$ Taken from \citet{ca2021} and references therein.}\\
\end{flushleft}
\end{table*}

By using different 1D, 2D and 3D methods {\citep[\textit{e.g.},][]{dr1988}}, applied to the distribution of the members {inside the caustics, these authors} searched for optical substructures in {the} cluster sample, supplementing their analysis with X-rays and radio literature data. 
{They found} that at least 70\% of the clusters in {their} sample present clear signs of substructuring, with 57\% being significantly substructured. 
The significance of the identified substructures in each cluster was estimated by the fraction of galaxies they contain respect to the total richness. 

The clusters were classified into five assembly state levels according to the presence ---or not--- and relative importance of substructures: 
unimodal systems, made up of a regular structure (U); low mass unimodal systems (L); systems with a primary structure and only low significance substructures (P);  significantly substructured systems with one main substructure (S); and multimodal conglomerates with more than one main substructure (M). 
While the L clusters are young poor galaxy systems, {representing evolutionary states prior to amalgamation processes but already relatively placidly evolved}, the U clusters are old and massive, which have probably grown by mergers and accretions and have already settled close to a relaxation state. 
P clusters are also old and massive, but still present signs of recent accretions; since these accretions are minor, the relaxation state of the cluster is almost unaffected. Finally, S and M clusters are systems during merging processes, the difference being if these mergers are minor or major, respectively. 

The benefit of using this sample lies in the {availability of this discrete classification of} 
%fact that each cluster is classified into one of the levels of 
{gravitational} assembly states 
%(U, P, S, M and L, 
(see Column 2 of Table \ref{tab:prop} {below}). 
%described in section \ref{sec:assembly}. 
This will serve to establish correlations between 
%the values thrown by the entropy-based estimator 
{$H_Z$ estimator} 
applied to the ensemble of galaxies of each cluster and 
the evolutionary state {obtained from direct} 
%estimated by 
observational methods. 

The {observational} cluster sample is reported in Table \ref{tab:init}: Column 1 shows the name of clusters; Columns 2, 3 and 4 present their right ascension, declination and mean redshift coordinates, respectively; Column 5 presents the number of presumably {virialized} sampled galaxies belonging to the system (inside the caustics and the aperture 1.3$\times r_{200}$); Column 6 presents the radius $r_{500}$ at which the mean interior overdensity is 500 times the critical density at the corresponding redshift; and Column 7 presents the X-ray temperature of clusters. 

For each observed cluster, 
%in the Top70 observational sample, 
the virial mass was estimated using (\ref{M_vir}) and the virial radius by 
\begin{equation}\label{R_v}
R_{\mathrm{vir}}^3=\frac{3\mathcal{M}_{\mathrm{vir}}}{4\pi\rho_{\mathrm{vir}}}=\frac{\alpha\sigma_{LOS}^2 R_{\mathrm{p}}}{18\pi H^2(z)}.
\end{equation}
where $\rho_{\mathrm{vir}}=18\pi^2[3H^2(z)/8\pi G]$ is the virialization density assuming a spherical model for nonlinear collapse \citep[\textit{e.g.},][]{bry1998}. 
In all calculations we used $\alpha = 2.5$, assuming a weak anisotropy \citep[\textit{e.g.},][]{tl2015,ka2020}.
{The} line-of-sight velocity dispersions were computed using the Tukey's biweight robust estimator \citep{br1990}. The results of $\sigma_{LOS}$, $\mathcal{M}_\mathrm{vir}$ and $R_\mathrm{vir}$ are shown, respectively, in Columns 8, 9 and 10 of Table \ref{tab:init}.  

% Subsection 4.2
%%%%%%%%%%%%%%%%%%%%%%%%%%%%%%%%%%%%%%%%%%%%%%%%%%%%%%%%%
\subsection{Simulation data} \label{ssec:ilus}

In addition, we use data from the IllustrisTNG simulation \citep[\textit{e.g.},][]{sp2018,nel2019}, a set of cosmological simulations that {assume initial conditions consistent with the \citet{pc2016} results and take into account magnetohydrodynamical effects}. In particular, we use the TNG300-1 data cube that contemplates the largest volume {allowing} the study of the distribution of galaxies and massive objects such as clusters. We take halos (equivalent to galaxy clusters) from the TNG300-1 simulation at redshift $z=0$,\footnote{The positions are given in rectangular coordinates of the form $(x,y,z) \, ah^{-1}$ kpc, where $a$ is the cosmological scale factor, being $a=1$ in $z=0$, and $h=H_0 /(100$ km s$^{-1}$ Mpc$^{-1})$, for which a value of $h=0.7$ was assumed here.} which has a resolution of $2500^3$ dark matter particles and $2500^3$ baryonic matter particles. We sampled 248 halos with masses greater than $10^{14} h^{-1} \mathcal{M}_\odot$ with 370,119 member subhalos, of which 366,940 are classified as galaxies. Information related to the peculiar velocity, mass and position of the center of each halo and subhalo was extracted.  

Each halo in the simulation contains, on average, 1,300 subhalos within its virial radius, greatly outnumbering the tracers {(galaxies)} in our observational sample of clusters. This large difference is due to the number of low-mass systems (dwarf galaxies {subhalos}) entering the count of TNG300-1; these galaxies are hard to detect in {real} clusters due to their low luminosity. Thus, to avoid statistical differences between the parameters estimated from {observational} and simulated data, we limit the selection of member subhalos by taking only those with mass greater than $2.0\times 10^{10} h^{-1} \mathcal{M}_\odot$. 
%{(see Fig. \ref{f:halos})}. 
For each halo, all member subhalos up to a distance of 1.3$\times r_{200}$ from the center (the particle with the least gravitational potential energy) were taken. {Also, $\sigma_{LOS}$, $\mathcal{M}_\mathrm{vir}$ and $R_\mathrm{vir}$ were calculated reproducing the observational procedure.}
%%%%%%%%%%%%%%%%%%%%%%%%%%%%%%%%%%%%%%%%%%%%%%%%%%%%%
\begin{figure*}[!ht]
  \begin{center}
       \includegraphics[trim={10cm 0cm 10cm 0cm},clip,scale=0.3]{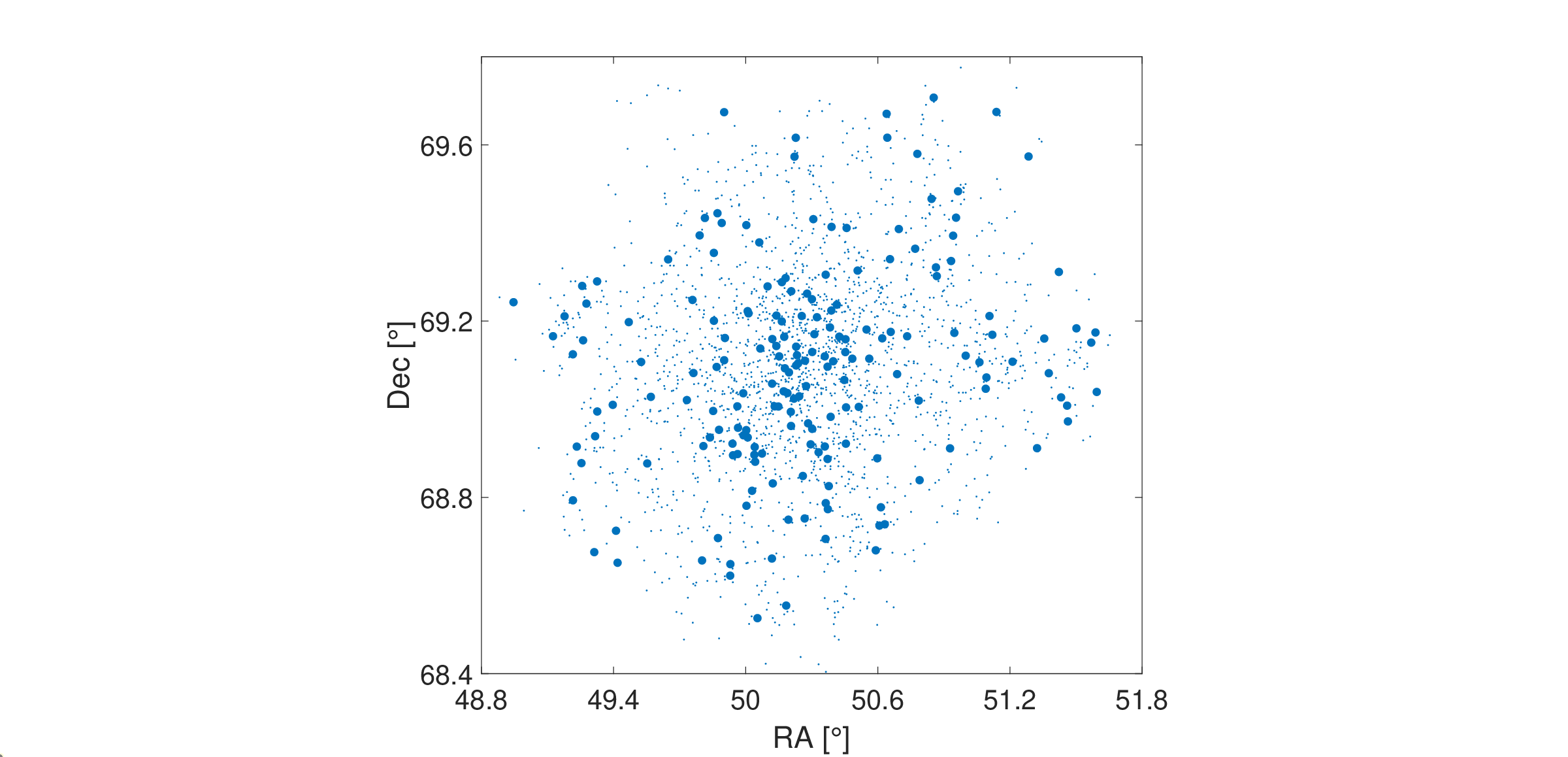}
       \includegraphics[trim={12cm 0cm 13cm 0cm},clip,scale=0.3]{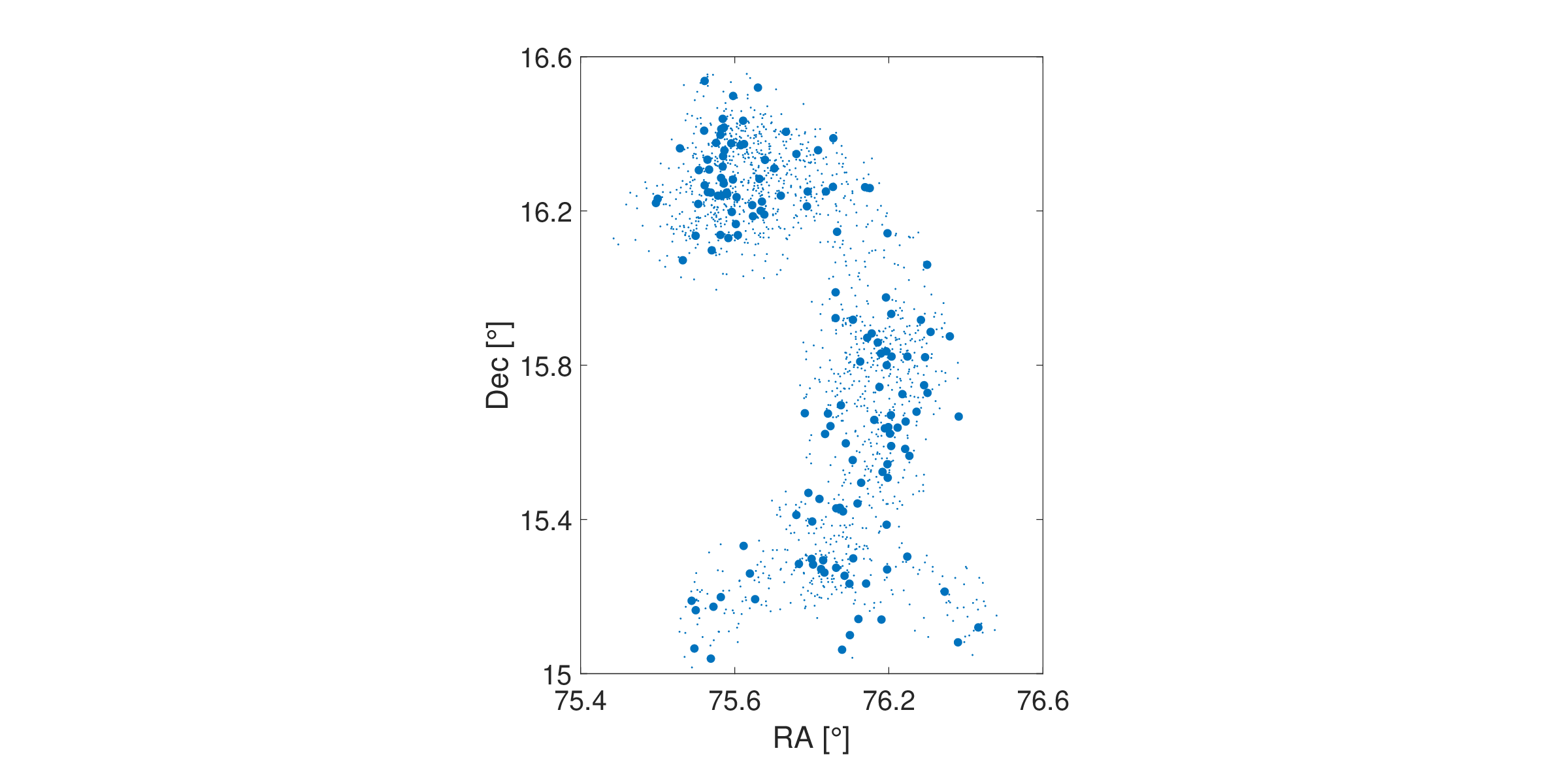}
\caption{Two examples of sampled TNG300 halos.  \textit{Left}: subhalo distribution for a high entropy halo (TNG-halo-34).  \textit{Right}: subhalo distribution for a low entropy halo (TNG-halo-87).  Each dot represents a member subhalo: the small dots are subhalos with masses less than $2.0\times 10^{10} \mathcal{M}_\odot$ while the big dots are the subhalos taken for our analysis.}
 \label{f:halos}
  \end{center}
\end{figure*}
%%%%%%%%%%%%%%%%%%%%%%%%%%%%%%%%%%%%%%%%%%%%%%%%%%%

Fig. \ref{f:halos} shows the projected distribution of member subhalos for two halos in the sample, one high-entropy (left) and one low-entropy (right). Note that, as expected, the distribution of subhalos is more random and homogeneous in a high entropy halo, while more substructured and elongated in a low entropy halo. This also happens in the observed clusters, {reinforcing} 
%confirming 
our hypothesis that evolutionary changes in galaxy systems progress in the direction in which the system dissolves substructures, becoming observationally more regular and homogeneous, with higher entropy values.

% Subsection 4.3
%%%%%%%%%%%%%%%%%%%%%%%%%%%%%%%%%%%%%%%%%%%%%%%%%%%%%%%%%
%\subsection{Entropy estimations in observed and simulated clusters}\label{subsec:est01}

\subsection{{Results on the assembly state of observed clusters}} \label{ssec:res01}

\begin{figure}[!ht]
  \begin{center}
   %<left> <lower> <right> <upper>
  \includegraphics[trim={1.5cm 0cm 2cm 0cm},clip,width=\columnwidth]{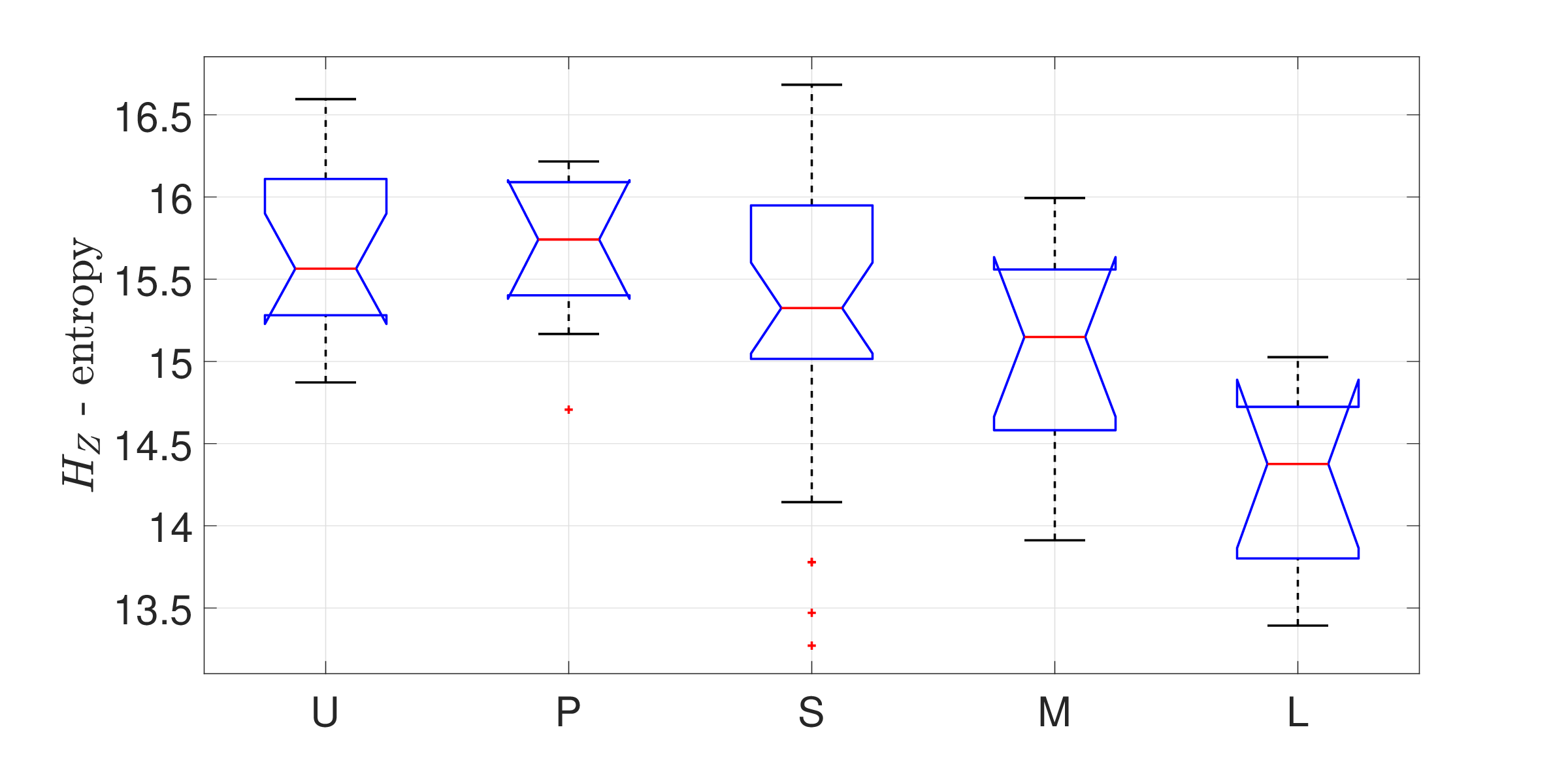}\\
\caption{Boxplots of $H_Z$-entropy for the five assembly classifications of clusters. The lower and upper extremes of each box are the 25th and 75th percentiles, respectively, while the central red line marks the median. Whiskers extend to the most extreme non-outlier data, while outliers are represented by red `+' symbols.}
 \label{f:boxHz}
  \end{center}
\end{figure}

To %better 
appreciate the correlation between the $H_Z$ entropy {(shown in Column 3 of Table \ref{tab:prop} {below})} and the gravitational assembly level ---and therefore with the evolutionary state--- of galaxy systems, we present in 
%the bottom panel of 
Fig. \ref{f:boxHz} the distributions of the $H_Z$ values with respect to the assembly state classes. 
{These are displayed} 
in the form of boxplots, {which} allow graphically describing the locality, dispersion and asymmetry of data classes (groups) of a quantitative variable through its quartiles \citep[\textit{e.g.},][]{hs2016}. The classes were ordered from more to less relaxed systems. {Although the box overlap does not allow one to unambiguously discriminate the class to which an arbitrary individual cluster should belong, the statistical} correlation is clear: the sequence U-P-S-M-L follows directly {the} decreasing {median values of} entropy. The only marginal difference case is between U and P systems, which is understandable if we consider that the primary systems (with only low significance substructures) are dynamically very similar to the unimodal ones ---both tend to be relatively massive and evolved systems. {This discussion will be resumed later.} 

%%%%%%%%%%%%%%%%%%%%%%%%%%%%%%%%%%%%%%%%%%%%%%%%%%%%%%%
%%%%%%%%%%%%%%%%%%%%%%%%%%%%%%%%%%%%%%%%%%%%%%%%%%%%%%%

%%%%%%%%%%%%%%%%%%%%%%%%%%%%%%%%%%%%%%%%%%%%%%%%%%%%%%%%%%%%%%%%%%
%TABLAS
\begin{table*}
\centering
\caption{Dynamical parameters for Top70  cluster sample.} \label{tab:prop} 
\resizebox{9cm}{!}{
%\hspace*{-3.5cm}
\begin{tabular}{lccrrrrcc}
\toprule %\hline \hline
%-----------------------------------------------------------------------
  \multicolumn{1}{c}{Name} &
  \multicolumn{1}{c}{$\mathcal{A}^\mathrm{a}$} & 
  \multicolumn{1}{c}{$H_Z$} &
  \multicolumn{1}{c}{$H_S$}&
  \multicolumn{1}{c}{$r'_c$}&
  \multicolumn{1}{c}{$\mathcal{P}_\mathrm{relax}$}&
  \multicolumn{1}{c}{$c_\mathrm{K}$}&
  \multicolumn{1}{c}{$c_\mathrm{NFW}$} &
  \multicolumn{1}{c}{$c_\mathrm{ICM}$} \\

  \multicolumn{1}{c}{} &
  \multicolumn{1}{c}{} &
  \multicolumn{1}{c}{} &
  \multicolumn{1}{c}{[nat]} &
  \multicolumn{1}{c}{[Mpc]} &
  \multicolumn{1}{c}{}&
  \multicolumn{1}{c}{}&
  \multicolumn{1}{c}{}&
  \multicolumn{1}{c}{} \\
  
   \multicolumn{1}{c}{(1)} &
  \multicolumn{1}{c}{(2)} &
  \multicolumn{1}{c}{(3)} &
  \multicolumn{1}{c}{(4)} & 
  \multicolumn{1}{c}{(5)} & 
  \multicolumn{1}{c}{(6)} &
  \multicolumn{1}{c}{(7)} &
  \multicolumn{1}{c}{(8)} &
  \multicolumn{1}{c}{(9)} \\
\hline
    A2798B & U & 15.54 & 10.85 & 0.28 & 0.808 &  7.85 & 3.91 & 2.33 \\
    A2801  & U & 15.34 & 10.60 & 0.29 & 0.794 &  3.94 & 1.22 &  ... \\
    A2804  & M & 14.58 & 10.16 & 0.22 & 0.668 &  3.41 & 1.55 &  ... \\
   A0085A  & S & 16.26 & 12.48 & 0.42 & 0.816 &  4.76 & 1.89 & 2.18 \\
    A2811B & S & 16.09 & 11.69 & 0.37 & 0.762 &  5.42 & 1.77 & 2.23 \\
    A0118  & S & 15.27 & 11.25 & 0.28 & 0.680 &  1.76 & 1.17 &  ... \\
    A0119  & S & 15.77 & 11.90 & 0.33 & 0.893 &  7.00 & 3.88 & 2.21 \\
    A0122  & U & 15.26 & 10.72 & 0.26 & 0.839 &  8.29 & 5.03 & 2.00 \\
    A0133A & S & 15.55 & 11.26 & 0.30 & 0.792 &  7.10 & 3.36 & 2.02 \\
    A2877-70  & S & 15.18 & 10.99 & 0.25 & 0.828 & 10.59 & 6.95 & 2.55 \\
AM0227-334 & L & 15.03 & 9.36  & 0.25 & 0.650 &  3.72 & 1.97 &  ... \\
    A3027A & S & 15.36 & 11.20 & 0.30 & 0.825 &  3.32 & 1.26 & 2.64 \\
    A0400  & S & 13.47 & 8.65  & 0.14 & 0.816 &  7.32 & 3.46 & 1.33 \\
    A0399  & U & 16.07 & 12.03 & 0.35 & 0.775 &  6.43 & 4.61 & 1.97 \\
    A0401  & U & 16.26 & 12.25 & 0.38 & 0.841 &  8.86 & 5.43 & 1.93 \\
    A3094A & U & 15.06 & 11.13 & 0.26 & 0.865 &  3.93 & 1.28 & 2.38 \\
    A3095  & L & 13.39 & 7.66  & 0.13 & 0.645 &  1.95 & 0.56 &  ... \\
    A3104  & S & 14.45 & 9.54  & 0.19 & 0.704 &  7.52 & 4.44 & 1.35 \\
    S0334  & L & 14.63 & 9.77  & 0.20 & 0.757 & 12.41 & 7.42 &  ... \\
    S0336  & L & 14.65 & 9.39  & 0.22 & 0.730 &  4.84 & 1.38 &  ... \\
    A3112B & S & 15.33 & 11.11 & 0.32 & 0.756 &  1.98 & 1.32 & 1.75 \\
    A0426A & P & 16.22 & 12.48 & 0.38 & 0.824 & 11.44 & 7.29 & 1.83 \\
    S0373  & S & 13.78 & 9.13  & 0.12 & 0.823 &  6.11 & 2.58 & 1.86 \\
    A3158  & S & 16.34 & 12.28 & 0.37 & 0.761 & 10.73 & 8.35 & 2.20 \\
    A0496  & S & 15.31 & 11.54 & 0.29 & 0.919 &  6.26 & 2.99 & 1.82 \\
    A0539  & S & 15.26 & 10.91 & 0.25 & 0.814 & 10.51 & 5.45 & 2.00 \\
    A3391  & U & 15.67 & 11.38 & 0.31 & 0.804 &  9.75 & 5.43 & 2.15 \\
    A3395  & M & 15.44 & 11.50 & 0.29 & 0.781 &  6.67 & 3.64 & 1.96 \\
    A0576  & S & 15.80 & 11.98 & 0.35 & 0.830 &  6.26 & 2.93 & 2.65 \\
    A0634  & L & 13.83 & 9.31  & 0.16 & 0.746 &  2.41 & 0.68 &  ... \\
    A0754  & M & 15.68 & 12.07 & 0.33 & 0.836 &  4.88 & 2.04 & 1.78 \\
    A1060  & P & 15.17 & 11.29 & 0.25 & 0.874 &  9.76 & 6.62 & 2.24 \\
    A1367  & M & 14.86 & 10.64 & 0.24 & 0.848 &  4.10 & 1.61 & 1.70 \\
    A3526A & P & 14.71 & 10.33 & 0.21 & 0.823 &  8.40 & 4.62 & 1.61 \\
    A3526B & S & 13.27 & 7.57  & 0.12 & 0.725 &  4.26 & 1.16 &  ... \\
    A3530  & S & 15.03 & 11.07 & 0.26 & 0.888 &  5.42 & 2.24 & 2.02 \\
    A1644  & P & 16.19 & 12.39 & 0.38 & 0.838 &  7.51 & 5.26 & 2.37 \\
    A3532  & S & 14.14 & 9.75  & 0.18 & 0.778 &  3.59 & 1.19 & 1.26 \\
    A1650  & U & 15.40 & 11.32 & 0.30 & 0.846 &  3.42 & 1.26 & 1.72 \\
    A1651  & P & 15.88 & 11.97 & 0.36 & 0.833 &  4.42 & 1.67 & 1.99 \\
    A1656  & S & 16.14 & 12.73 & 0.39 & 0.861 &  7.79 & 4.11 & 2.17 \\
    A3556  & M & 14.54 & 10.85 & 0.21 & 0.744 &  3.79 & 1.49 &  ... \\
    A1736A & S & 13.78 & 9.34  & 0.17 & 0.783 &  2.75 & 0.71 & 1.10 \\
    A1736B & S & 15.75 & 11.77 & 0.32 & 0.775 &  5.74 & 3.64 &  ... \\
    A3558  & P & 16.06 & 12.60 & 0.39 & 0.826 &  4.89 & 2.12 & 2.23 \\
SC1329-313 & L & 13.77 & 8.65  & 0.16 & 0.744 &  4.63 & 2.34 &  ... \\
    A3562  & U & 14.87 & 10.70 & 0.25 & 0.841 &  4.15 & 1.74 & 1.67 \\
    A1795  & U & 15.56 & 11.51 & 0.30 & 0.893 &  9.47 & 5.04 & 1.53 \\
    A2029  & U & 16.02 & 11.81 & 0.31 & 0.838 & 12.94 & 6.92 & 1.44 \\
    A2040B & S & 15.00 & 10.76 & 0.26 & 0.838 &  4.28 & 1.10 &  ... \\
\hline
\end{tabular}}
\end{table*}
%%%%%%%%%%%%%%%%
\begin{table*}[!ht]
\centering
{TABLE \ref{tab:prop} ---\textit{continued}} 
\resizebox{9cm}{!}{
%\hspace*{-3.5cm}
\begin{tabular}{lccrrrrcc}
\toprule %\hline \hline
%-----------------------------------------------------------------------
  \multicolumn{1}{c}{Name} &
  \multicolumn{1}{c}{$\mathcal{A}^\mathrm{a}$} & 
  \multicolumn{1}{c}{$H_Z$} &
  \multicolumn{1}{c}{$H_S$}&
  \multicolumn{1}{c}{$r'_c$}&
  \multicolumn{1}{c}{$\mathcal{P}_\mathrm{relax}$}&
  \multicolumn{1}{c}{$c_\mathrm{K}$}&
  \multicolumn{1}{c}{$c_\mathrm{NFW}$} &
  \multicolumn{1}{c}{$c_\mathrm{ICM}$} \\

  \multicolumn{1}{c}{} &
  \multicolumn{1}{c}{} &
  \multicolumn{1}{c}{} &
  \multicolumn{1}{c}{[nat]} &
  \multicolumn{1}{c}{[Mpc]} &
  \multicolumn{1}{c}{}&
  \multicolumn{1}{c}{}&
  \multicolumn{1}{c}{}&
  \multicolumn{1}{c}{} \\
  
   \multicolumn{1}{c}{(1)} &
  \multicolumn{1}{c}{(2)} &
  \multicolumn{1}{c}{(3)} &
  \multicolumn{1}{c}{(4)} & 
  \multicolumn{1}{c}{(5)} & 
  \multicolumn{1}{c}{(6)} &
  \multicolumn{1}{c}{(7)} &
  \multicolumn{1}{c}{(8)} &
  \multicolumn{1}{c}{(9)} \\
\hline
    A2052  & S & 15.08 & 10.98 & 0.25 & 0.850 &  6.82 & 3.69 & 1.69 \\
    MKW03S & U & 14.92 & 11.30 & 0.24 & 0.881 &  6.58 & 4.01 &  ... \\
    A2065  & U & 16.30 & 12.19 & 0.40 & 0.882 &  9.11 & 5.24 & 2.38 \\
    A2063A & P & 15.48 & 11.50 & 0.29 & 0.899 &  9.51 & 5.19 & 2.00 \\
    A2142  & P & 15.74 & 12.12 & 0.34 & 0.866 &  4.54 & 1.82 & 1.56 \\
    A2147  & M & 15.99 & 12.00 & 0.39 & 0.829 &  3.62 & 1.64 & 2.63 \\
    A2151  & M & 15.50 & 11.85 & 0.32 & 0.823 &  4.18 & 1.36 & 2.61 \\
    A2152  & M & 13.91 & 9.38  & 0.18 & 0.758 &  1.99 & 0.75 & 1.96 \\
    A2197  & M & 14.76 & 10.65 & 0.25 & 0.699 &  1.59 & 1.06 & 3.12 \\
    A2199  & P & 15.53 & 12.02 & 0.30 & 0.884 &  7.41 & 4.23 & 1.89 \\
    A2204A & S & 16.51 & 11.16 & 0.41 & 0.769 &  3.92 & 1.72 & 1.84 \\
    A2244  & U & 16.60 & 12.33 & 0.41 & 0.821 & 10.46 & 7.81 & 2.25 \\
    A2256  & S & 16.68 & 12.47 & 0.43 & 0.607 & 10.87 & 8.49 & 2.39 \\
    A2255  & S & 16.20 & 12.31 & 0.39 & 0.789 &  6.54 & 3.42 & 2.31 \\
    A3716  & M & 15.56 & 11.56 & 0.30 & 0.769 &  5.53 & 3.17 &  ... \\
    S0906  & L & 14.12 & 9.30  & 0.18 & 0.793 &  2.94 & 0.74 &  ... \\
    A4012A & L & 14.80 & 10.13 & 0.23 & 0.834 &  4.26 & 1.63 &  ... \\
    A2634  & S & 15.32 & 11.60 & 0.28 & 0.861 &  6.69 & 3.81 & 2.38 \\
A4038A-49 & S & 15.45 & 11.35 & 0.28 & 0.820 &  9.73 & 5.12 & 2.01 \\
    A2670  & U & 16.12 & 11.92 & 0.33 & 0.859 & 10.24 & 6.01 & 2.29 \\
\hline
\end{tabular}}
\begin{flushleft}  
\footnotesize{$^\mathrm{a}$ Gravitational assembly classes from \citet{ca2021}.}\\
\end{flushleft}
\end{table*}
%%%%%%%%%%%%%%%%%%%%%%%%%%%%%%%%%%%%%%%%%%%%%%%%%%%%%

%%%%%%%%%%%%%%%%%%%%%%%%%%%%%%%%%%%%%%%%%%%%%%%%%%%%%%%%%
% Section 5
%%%%%%%%%%%%%%%%%%%%%%%%%%%%%%%%%%%%%%%%%%%%%%%%%%%%%%%%%
\section{Testing statistically the $H_Z$-entropy estimator} \label{sec:shannon}

% Subsection 5.1
%%%%%%%%%%%%%%%%%%%%%%%%%%%%%%%%%%%%%%%%%%%%%%%%%%%%%%%%%
\subsection{{Shannon entropy of galaxy distributions in phase-spaces}} \label{sec:H_S}

{Another way we use to evaluate  %validate 
the entropy estimates that $H_Z$ can provide  
%an alternative method is needed 
is by using a method} that does not take into account equilibrium assumptions, but allows {to characterize the internal states of a galaxy system from the} raw distribution of its member galaxies. {In information theory, for example,} entropy is a measure of the uncertainty of a random variable \citep[or source of information, \textit{e.g.},][]{sh1948,cov2006}.
%, {and this approach is useful to meet the above goal}. 
If $X$ is a discrete random variable with possible observable values $x\in \mathcal{X}$, which occur with probability $p(x)=\mathrm{Pr}\left\lbrace X=x\right\rbrace$, the Shannon (or information) entropy of $X$ is defined as
\begin{equation}\label{Shan}
H_S(X)\equiv-\sum_{x\in \mathcal{X}} p(x)\log_{\lambda}{p(x)},
\end{equation}    
where the sum is performed over all possible values of the variable, and the base $\lambda$ of the logarithm is chosen according to the entropy `units' to use (\emph{nats}, $\lambda=e$; \emph{bits}, $\lambda=2$; \emph{bans}, $\lambda=10$). One of the criteria used by \citet{sh1948} to define $H_S$ ensures that it increases as the possible values of $X$ begin to appear with equal frequency, taking higher values when they become equiprobable, \textit{i.e.}, when there are no special configurations in the data distribution that provide more information, increasing the uncertainty. 

\begin{figure*}[!ht]
\centering
%<left> <lower> <right> <upper>
\includegraphics[trim={5cm 0cm 2.5cm 1cm},clip,width=0.9\textwidth]{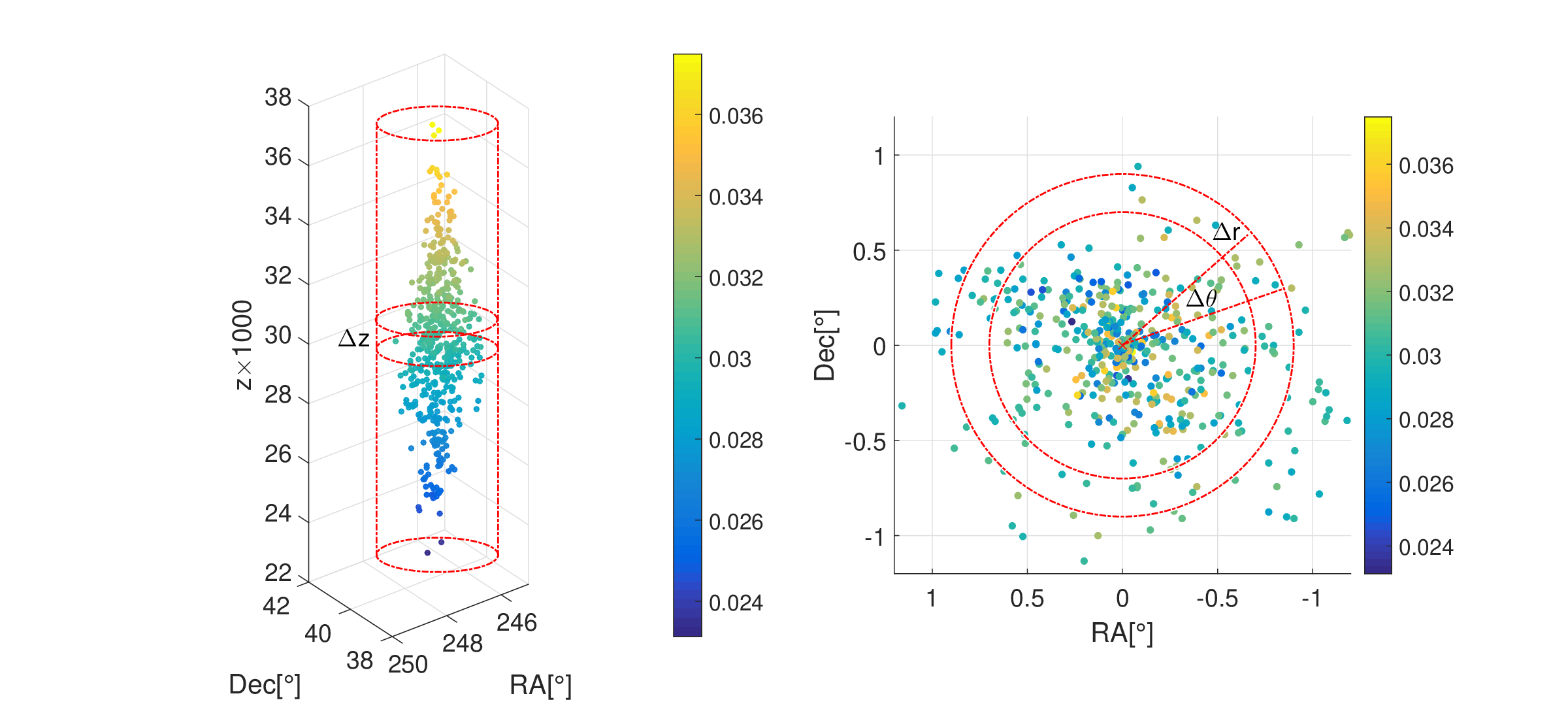}
 \caption{Sample of galaxies for A2199 cluster in \citet{ca2021}. Color bars represent the redshift ($z$) distribution. \textit{Left panel}: the cylinder along the line-of-sight 
direction. \textit{Right panel}: projected distribution in the sky plane with RA 
and Dec coordinates transported to the origin (the FRG, see the text).}
 \label{f:cyl}
\end{figure*}

{Now,} {the raw coordinates of the galaxies in a cluster, \textit{i.e.} the observable set of triples $($RA$,$Dec$,z)$ of right ascension, declination and redshift, are distributed within a solid angle, which can be approximated by a cylinder with a circular base in the plane of sky and depth along the line-of-sight (see, Fig. \ref{f:cyl}, left panel). Inside the cylinder, the position of each galaxy can be expressed in the form $x=(r,\theta,z)$, where $r$ is its projected distance from the cluster center, $\theta$ its ---azimuthal--- angle with respect to the local north direction in the projected sky distribution (see, Fig. \ref{f:cyl}, right panel), and the redshift $z$ a measure of its radial velocity.} {In fact,} the cylinder of $x$-coordinates {(two spatial and one velocity) may be considered} a ---projected--- phase-space for the {galaxy ensemble, and the distribution of the variables $(r,\theta,z)$ inside it depend on the dynamical state of the galaxy cluster.}  

%%%%%%%%%%%%%%%%%%%%%%%%%%%%%%%%%%%%%%%%%%%%%%%%%%%%
%%%%%%%%%%%%%%%%%%%%%%%%%%%%%%%%%%%%%%%%%%%%%%%%%%%%

{By considering} the cylinder of sampled galaxies as a source of information {for the $x$-distribution}, the probability of finding a galaxy in the neighborhood of position $x$ (\textit{i.e.}, the probability that the position random variable $X$ takes a value very close to $x$ inside the cylinder) can be approximated as $p(x)\simeq \bar{f}_{r \theta z}(x)\Delta x$, where $\bar{f}_{r\theta z}$ must be the observed ---or empirical--- joint probability density function (PDF) that represents the actual distribution of the galaxy ensemble {in the variable $x$}. Replacing this in (\ref{Shan}), one can compute the {Shannon} entropy of the galaxy distribution for a cluster {in the form} 
\begin{equation} \label{Sh_glx}
H_S=-\sum_{x\in \mathcal{X}}^n \left[\bar{f}_{r\theta z}(x)\Delta x \right] \ln{\left\lbrace \bar{f}_{r\theta z}(x)\Delta x \right\rbrace},  
\end{equation}
where $\lambda=e$ has been chosen, and the sum is performed over all the $n$ discrete partitions $\Delta x$ made to the domain $\mathcal{X}=[0,R_{\mathrm{vir}}]\times[0,360]\times[z_{\mathrm{min}},z_{\mathrm{max}}]$ of the observed 
joint PDF. Here, $z_{\mathrm{min}}$ and $z_{\mathrm{max}}$ are the minimum and maximum values of redshift in the sample of galaxies of each cluster. 

{Shannon entropy is a well-established measure for quantifying disorder or uncertainty in a system \citep[\textit{e.g.},][]{cov2006}. In this context, $H_S$ provides a measure of the degree of randomness in the distribution of galaxies in the phase-space of a cluster (or any other galaxy system). By calculating this entropy, we are evaluating the information contained in the galaxy ensemble and how galaxies are distributed in different regions of phase-space.} 
{Close to equilibrium, the memory (information) of the initial conditions of clusters formation is lost \citep[\textit{e.g.},][]{bt2008,am2009}}. 
The 
%more
general character {of $H_S$} comes from the fact that it is not restricted only to thermodynamic variables, but to any type of data $X$ that contains information about the state of a system, {so it can be used to study characteristics of non-equilibrium systems as they evolve.} In addition, it does not consider the microstates of a system as equiprobable, keeping a certain relationship in mathematical form and meaning with the Gibbs entropy \citep[\textit{e.g.},][]{ja1957}. \\

% Subsection 5.2
%%%%%%%%%%%%%%%%%%%%%%%%%%%%%%%%%%%%%%%%%%%%%%%%%%%%%%%%%
\subsection{{Shannon entropy estimations}}\label{subsec:est02}

\begin{figure}[!ht]
  \centering
  %<left> <lower> <right> <upper>
    \includegraphics[trim={1cm 0cm 3cm 0cm},clip,width=\columnwidth]{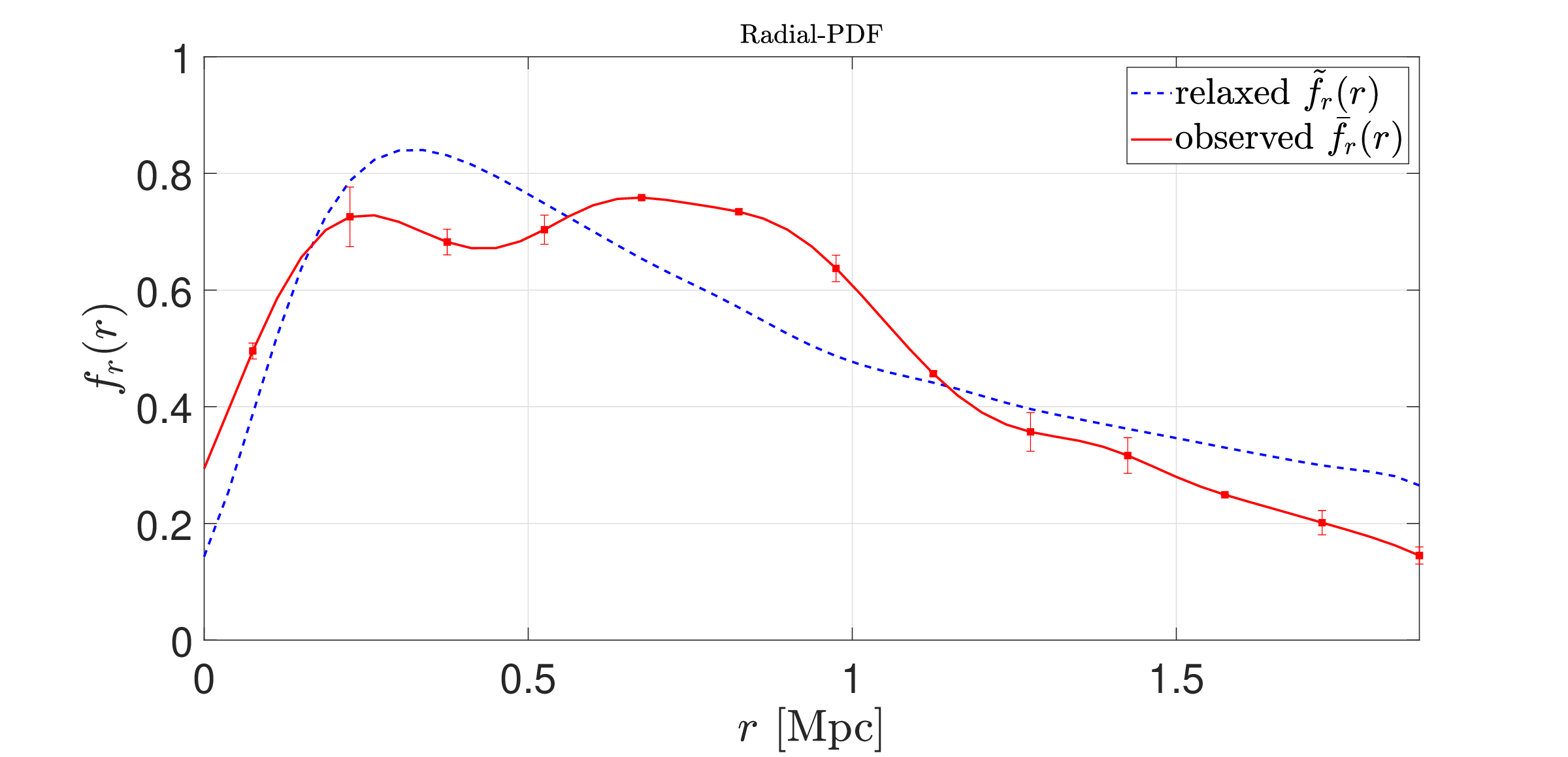}\\
    \includegraphics[trim={1cm 0cm 3cm 0cm},clip,width=\columnwidth]{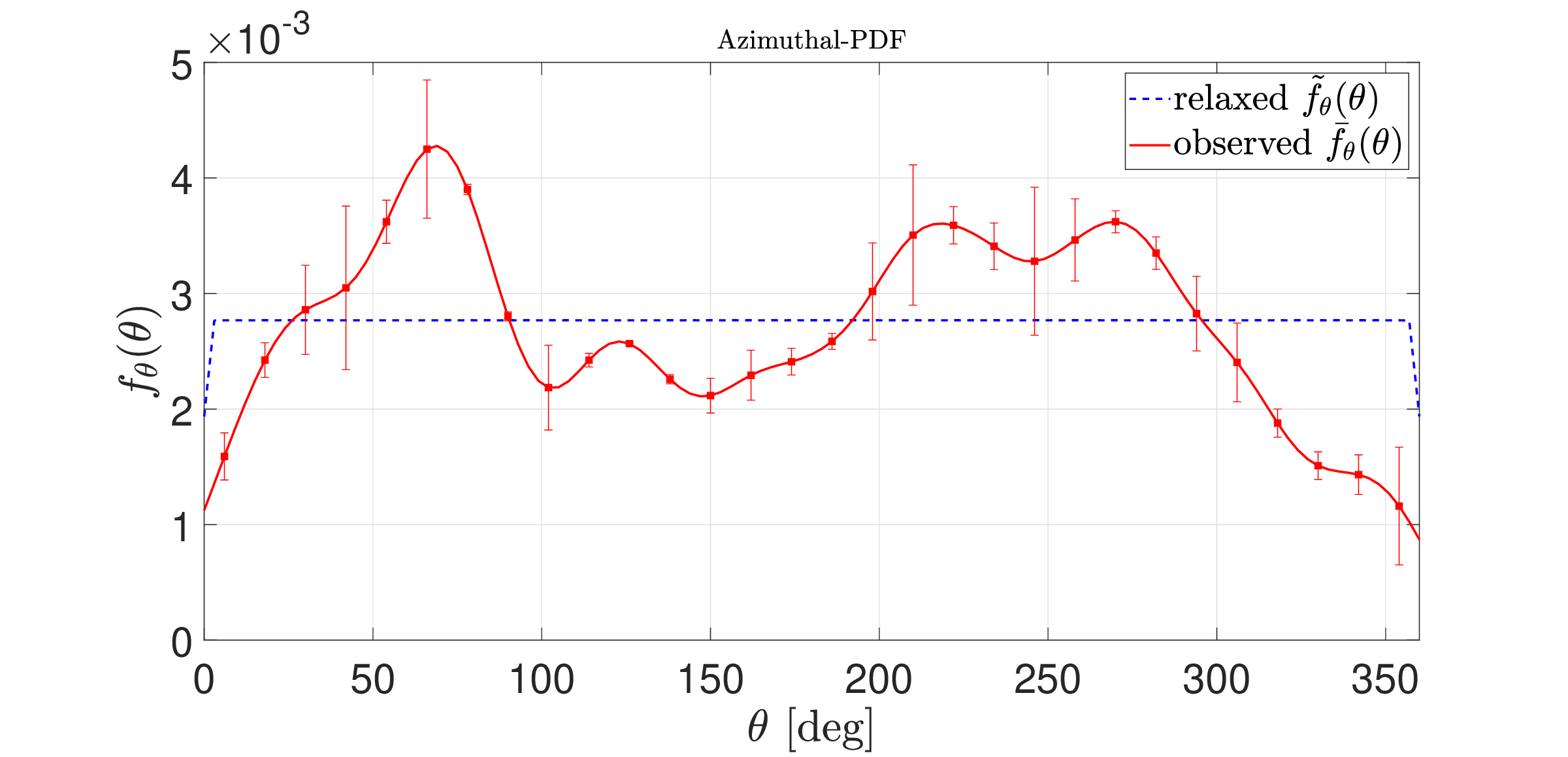}\\
    \includegraphics[trim={1cm 0cm 3cm 0cm},clip,width=\columnwidth]{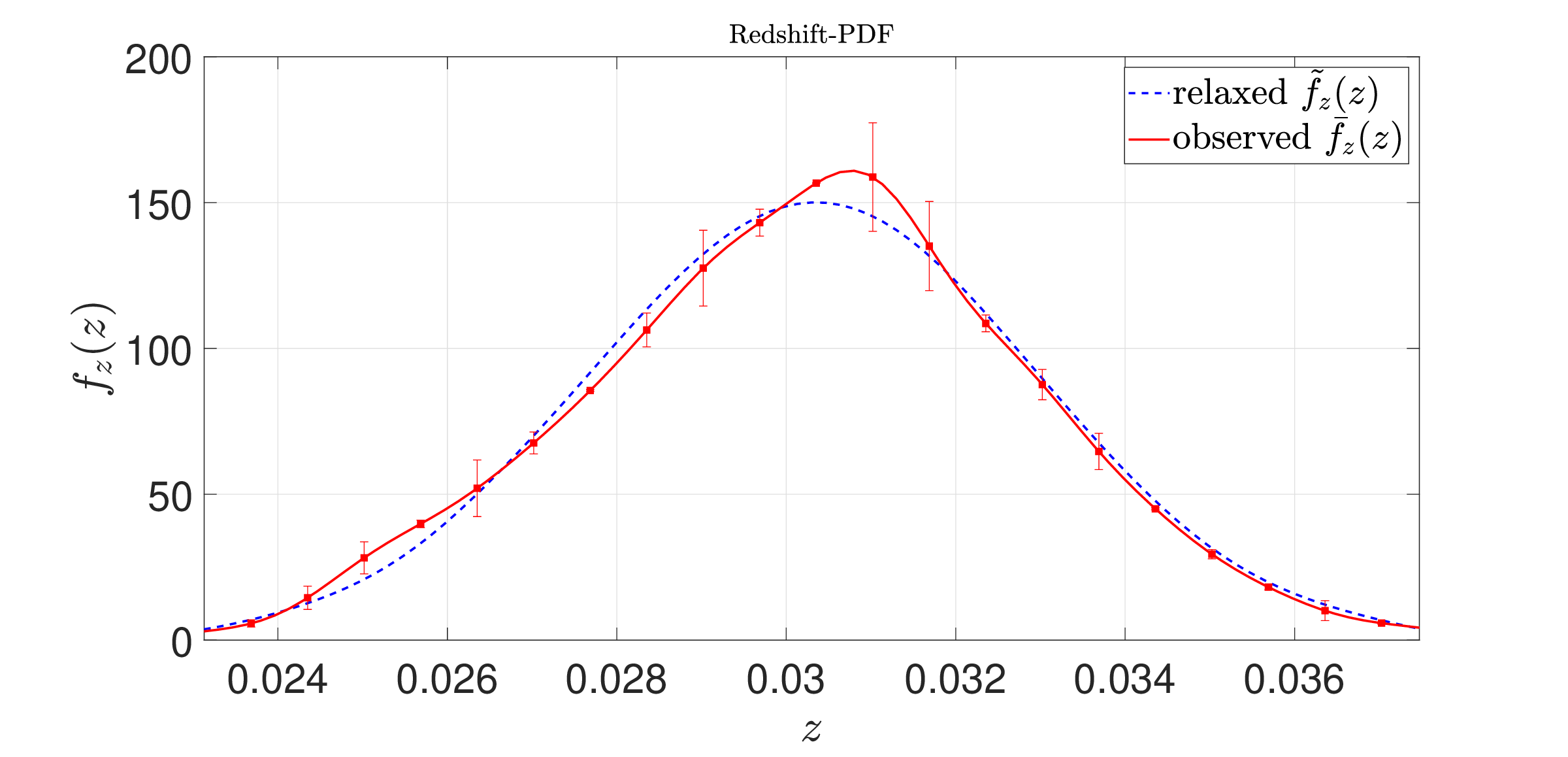}
 \caption{Probability density functions for the radial-$r$, azimuthal-$\theta$ and redshift-$z$ distributions of member galaxies in the A2199 cluster. The red solid lines represent the \emph{observed PDFs} obtained by smoothing (normal kernel) from observational data histograms, while dashed blue lines represent the \emph{relaxed PDFs} of equilibrium distributions. The error bars represent the difference between the height of the smoothed curve and the height of the corresponding bin in each 1D-histogram.}
 \label{f:pdf}
\end{figure}

{To calculate the Shannon entropy, we first established the observed joint PDF $\bar{f}_{r \theta z}(x)$ of each cluster in two different ways. First,} by counting galaxies independently in bins of width $\Delta r$, $\Delta \theta$ and $\Delta z$ in the radial-$r$, azimuthal-$\theta$ and redshift-$z$ directions, respectively, {we constructed} 1D-histograms that describe the observed distribution of that variables in the galaxy ensemble. {Then,} by a smoothing technique, the %{respective} 
\emph{observed PDFs} $\bar{f}_r(r)$, $\bar{f}_{\theta}(\theta)$ and $\bar{f}_z(z)$ {were obtained for the variables} from their respective normalized histograms (see, red solid lines on graphs in Fig. \ref{f:pdf}). 
{For this,} {we used a kernel density estimator {that allows a} non-parametric fit of PDFs of random variables, {adapting} directly to the data \citep[\textit{e.g.},][]{hs2016}. This method is particularly useful when the actual distribution of a data set is unknown, as is the case for the galaxy distributions inside the (phase-space) cylinders. A standard Gaussian smoothing kernel was used with bins of widths $\Delta r=0.15$ Mpc, $\Delta\theta=12^\circ$, $c\Delta z=200$ km s$^{-1}$ and supports in the intervals $[0,R_{\mathrm{vir}}]$, $[0,360]$ and $[z_{\mathrm{min}},z_{\mathrm{max}}]$ for the $\bar{f}_r$, $\bar{f}_{\theta}$ and $\bar{f}_z$ distributions, respectively.  Finally, we took $\bar{f}_{r\theta z}(r,\theta,z) = \bar{f}_r(r)\bar{f}_{\theta}(\theta)\bar{f}_z(z)$ assuming statistical independence of the variables $r$, $\theta$ and $z$ in the galaxy distributions.} 
%%%%%%%%%%%%%%%%%%%%%%%%%%%%%%%%%%%%%%%%%%%%%%%%%%%%
%%%%%%%%%%%%%%%%%%%%%%%%%%%%%%%%%%%%%%%%%%%%%%%%%%%%

{On the other hand,} {as is evident in the right panel of Fig. \ref{f:cyl}, galaxies with the highest and lowest redshifts prefer the center of the projected distribution of the cluster. This correlation between the variables $r$ and $z$ is physically justified since the galaxies acquire a greater speed during their transit through the central regions of the clusters (\textit{i.e.}, where the gravitational potential well is the deepest). Projection effects can also occur, especially for halo galaxies that are located in front or behind the core and close to the line-of-sight. No significant {correlation was} detected between the variables $r$ or $z$ with $\theta$ {instead}.} 
Thus, to {construct} the observed joint PDF {in the second way we only took} into account the correlation between $r$ and $z$, such that $\bar{f}'_{r \theta z}(r,\theta,z)=\bar{f}_{rz}(r,z)\bar{f}_{\theta}(\theta)$, where the $\bar{f}_{rz}$ functions were {obtained} by counting galaxies in {bi-dimensional} bins of ``area'' $\Delta r \Delta z$ and smoothed by a multivariate Gaussian surface kernel, using the same bin width values as above. 

Two statistical tests, Kolmogorov-Smirnov and Rank-Sum, showed no significant differences between both methods, confirming the null hypothesis that the PDFs $\bar{f}_{r \theta z}$ and $\bar{f}'_{r \theta z}$ represent the same distribution with a confidence level of {95\%} in both tests. This is probably because the fraction of galaxies that present a strong $rz$-correlation is very small, implying a low statistical weight. {Then,} for simplicity we choose the first option (\textit{i.e.}, {that} with the three variables assumed to be statistically independent{, see Fig. \ref{f:pdf}}) {to compute the probability values}
\begin{equation}
p(x)\simeq\bar{f}_{r\theta z}(x)\Delta x=[\bar{f}_r(r)\Delta r][\bar{f}_{\theta}(\theta)\Delta\theta][\bar{f}_z(z)\Delta z],    
\end{equation}
{varying one of the variables in its respective domain (\textit{e.g.}, $r\in[0,R_{\mathrm{vir}}]$, $\theta\in[0,360]$ and $z\in[z_{\mathrm{min}},z_{\mathrm{max}}]$), while keeping the other two constant.} {Thus, $p(x)$ was obtained in about 1,000 positions $x=(r,\theta,z)$} {inside the data cylinder of each real (and simulated) cluster {and} these values were used to compute the respective $H_S$-entropies by (\ref{Sh_glx}) for the galaxy ensembles. The results for \emph{Top70} clusters are shown in Column 4 of Table \ref{tab:prop}.}

% Subsection 5.3
%%%%%%%%%%%%%%%%%%%%%%%%%%%%%%%%%%%%%%%%%%%%%%%%%%%%%%%%%
\subsection{{Results on the Shannon entropy for observed and simulated clusters}} \label{subsec:res03}

\begin{figure}[!t]
  \begin{center}
   %<left> <lower> <right> <upper>
  \includegraphics[trim={2.5cm 0cm 3cm 0cm},clip,width=\columnwidth]{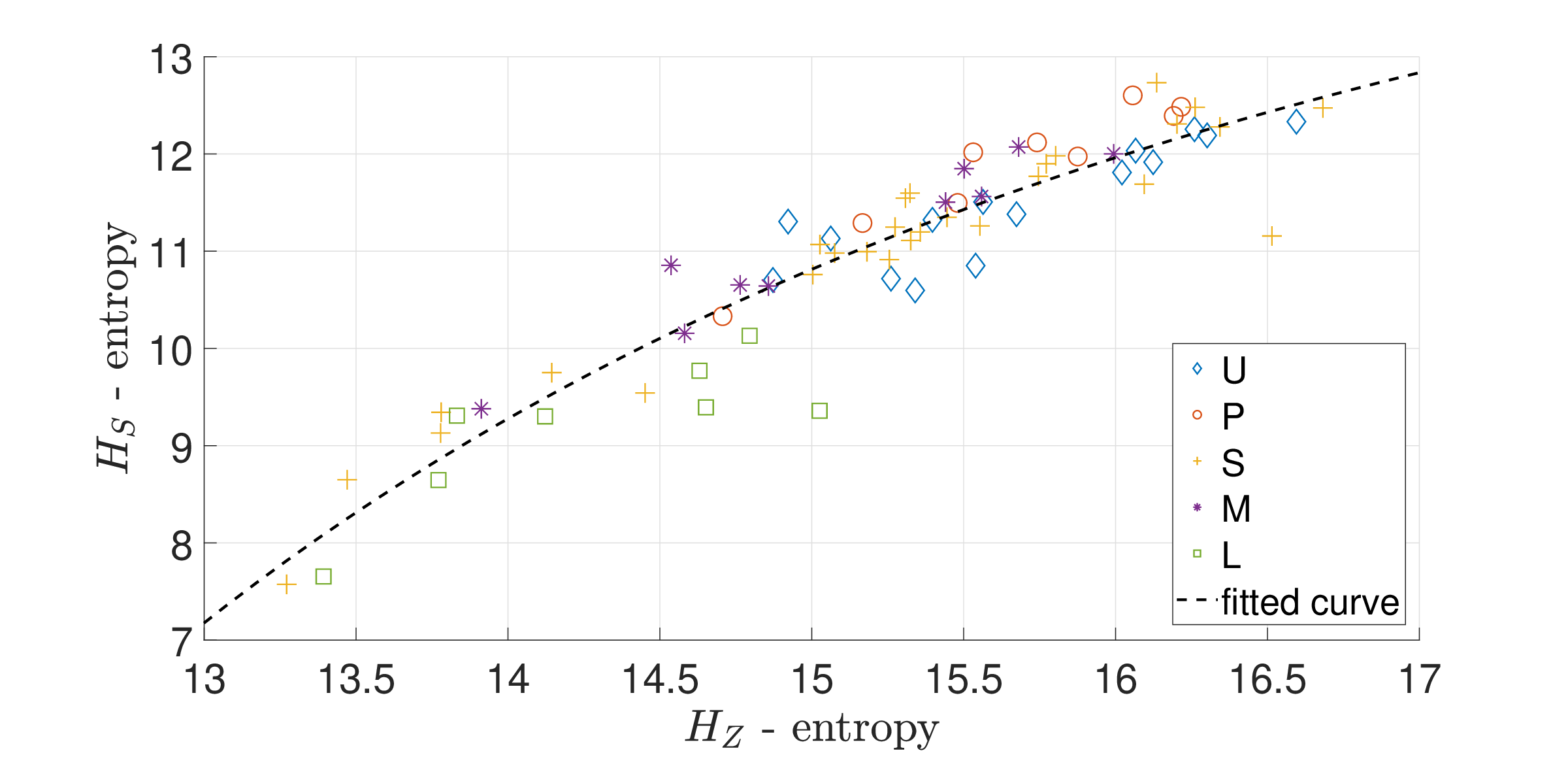}\\
\caption{Scatter plot of $H_Z$ \textit{vs.} $H_S$ made with the entropy values estimated for the \emph{Top70} clusters. The symbols and color scale of the points represents the U-P-S-M-L assembly level classification of clusters performed by \citet{ca2021}. The dashed line represents the best ---power law--- fit with $\mathcal{R}^2=0.886$.}
 \label{f:HzHs}
  \end{center}
\end{figure}

{The relation between the $H_Z$-entropy estimator,} proposed in the present work, {and the Shannon entropy $H_S$ of the real galaxy distributions} can be seen in Fig. \ref{f:HzHs}, where the dashed line {represents} the best fit curve, with a coefficient of determination \citep[$\mathcal{R}^2$, \textit{e.g.},][]{hs2016} {of 0.886}. 
This figure shows {a high degree of} correlation (almost linear) between $H_Z$ and $H_S$ entropies that, {when measured by the Pearson} and Spearman coefficients \citep[\textit{e.g.},][]{gc2003,hs2016}, {gives the values 0.932 and 0.922, respectively.} 
Below (see, section \ref{sec:disc}), we offer what we believe to be a possible explanation for such an explicit correlation. 
{In addition, the figure} also shows the association of these $H_Z$ and $H_S$ entropy values with the assembly state of the clusters (represented by the U-P-S-M-L classes). Although the correlation is not evident, it is noticeable that all U and P clusters are placed in the locus of points with higher values of both $H_Z$ and $H_S$, while all L are located in the region with lower values of these entropies. 

\begin{figure}[!t]
  \centering
  %<left> <lower> <right> <upper>
    \includegraphics[trim={1cm 0cm 3cm 0cm},clip,width=\columnwidth]{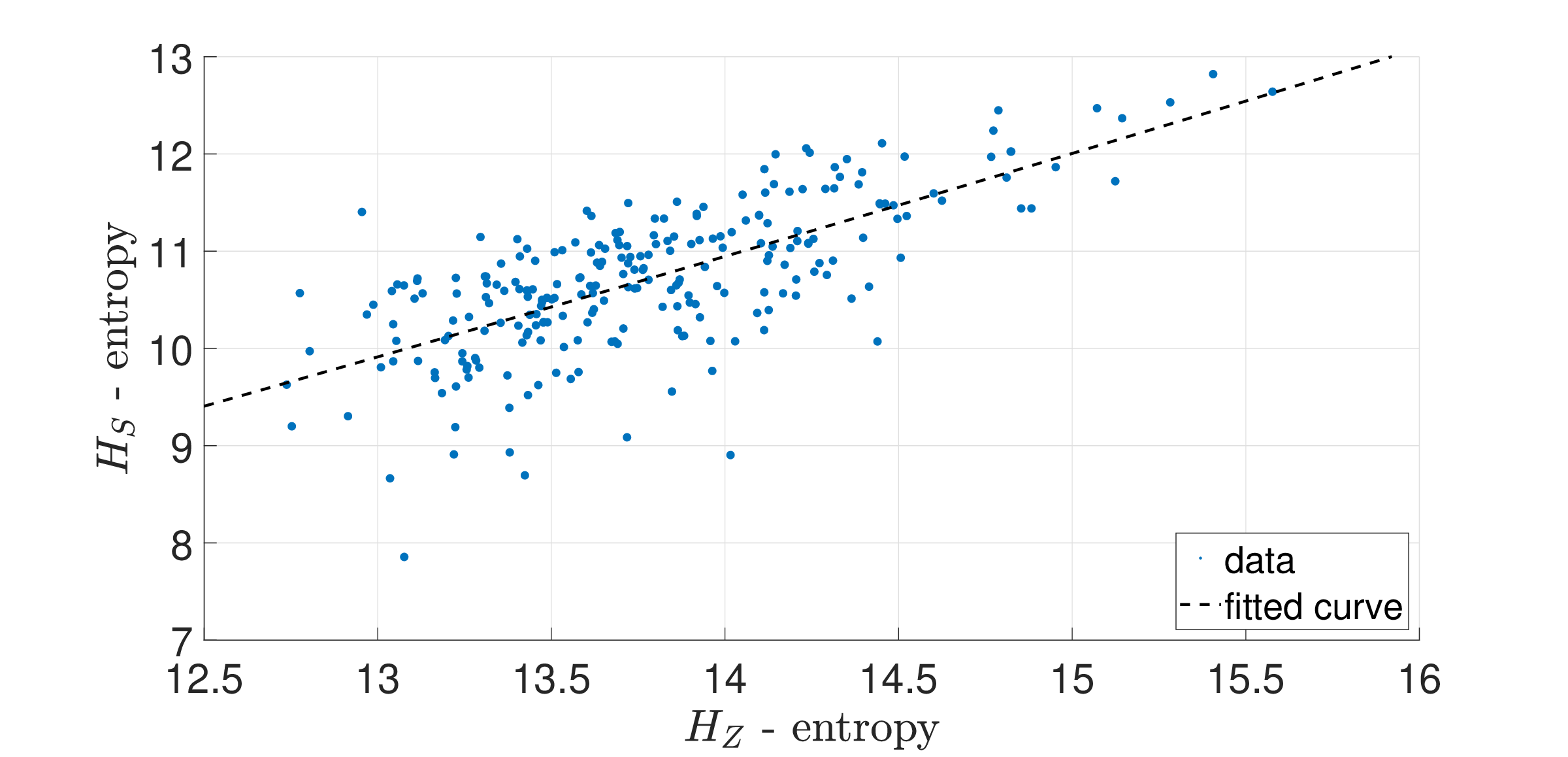} \\
 \caption{Scatterplot of $H_S$ \textit{vs.} $H_Z$ entropies made with the data of the complete TNG300 sample. The dashed line represents the best power law fit with $\mathcal{R}^2=0.505$.}
 \label{f:HzHsTNG}
\end{figure}

A procedure similar to that performed with \emph{Top70} clusters was applied to the TNG300 sample {to compare the $H_Z$ and $H_S$ entropies estimated for the simulated} halos. Fig. \ref{f:HzHsTNG} {shows a significant correlation, with Pearson and Spearman coefficients of 0.709 and 0.678, respectively, between the {dynamical} and Shannon entropies,} which reveal to be very similar to that of real clusters. {We did not carry out a classification of simulated clusters in their different assembly levels ($\cal{A}$) that allows us to analyze the distribution of $H_Z$ in each class, as in the case of real clusters. We hope to do this in future work.} \\

%%%%%%%%%%%%%%%%%%%%%%%%%%%%%%%%%%%%%%%%%%%%%%%%%%%%%%%%%
% Section 6
%%%%%%%%%%%%%%%%%%%%%%%%%%%%%%%%%%%%%%%%%%%%%%%%%%%%%%%%%
\section{Other validations for $H_Z$} \label{sec:valid}

% Subsection 6.1
%%%%%%%%%%%%%%%%%%%%%%%%%%%%%%%%%%%%%%%%%%%%%%%%%%%%%%%%%
\subsection{{The relaxation probability of galaxy systems}}\label{subsec:res3}

{Several} studies reveal that clusters {close} to dynamical equilibrium have 
distributions of galaxies whose 
radial density and {LOS} velocity {profiles} tend to ones that can be represented by specific mathematical functions \citep[\textit{e.g.},][]{,sas1984,sa1988,ad1998,sam2014}. 
{The above is also true for the ICM entropy profiles in X-ray observations \citep[\textit{e.g.},][]{voi2005} or for the distribution of DM subhalos in cosmological simulations \citep[\textit{e.g.},][]{nfw1996}.} {Thus, we consider} that it is {also} possible to {characterize} the evolutionary state of a galaxy system by measuring how far the {currently \emph{observed PDFs} are} 
%of any of its dynamical parameters is 
from {their} expected equilibrium {functional} shapes. For this, we need to {use (or choose)} a consistent reference {equilibrium} model that describes the distribution 
%of some observable dynamical parameter 
when the galaxy ensemble is relaxed, {as well as} a metric that allows us to estimate the distance of the observed distribution from that of the {equilibrium} model. 

We limit ourselves to {a} simple reference {equilibrium} model, considering clusters {represented} by a spherical distribution of galaxies, with a homogeneous core-halo spatial configuration and {an isotropic} velocity distribution without net angular momentum. 

{In principle, the radial distribution of galaxies can be well described \citep[see,][]{ad1998} by different single popular mass profiles proposed for clusters in the literature, both \emph{core-} \citep[\textit{e.g.}, King, Einasto; respectively][]{ki1962,ei1965} and \emph{cuspy-} \citep[\textit{e.g.}, Hernquist, NFW, respectively][]{her90,nfw1996}
dominated ones.} 
{Nonetheless, there has been a tendency, especially for fitting DM halos, to use more complex functions, with a larger number of free parameters \citep[\textit{e.g.},][]{Deh93,Fie20,Die23} --although they are essentially double and/or truncated power laws--, in order to better accommodate the inner and outer slopes.} 
{On the other hand, since \emph{core}-dominated profiles usually perform slightly better for real clusters \citep[\textit{e.g.},][]{sa1988,ad1998,GaM22} ---as commented before, the core-halo structure is the one expected for self-gravitating systems at equilibrium--- and considering the simplest as possible analytical form, we choose the King density profile at this point.} 
{Such profile} can be approximated analytically {by} the form
\begin{equation}\label{K_3d}
\rho(r)=\rho_0\left[1+\left(\frac{r}{r_c} \right)^2 \right]^{-\gamma},
\end{equation}  
suggesting a finite central density $\rho_0=\rho(0)$ and the existence of a core of radius $r_c$. The parameters $r_c$, $\rho_0$ and $\gamma$ {can} be determined by the best fit of the model (\ref{K_3d}) to the spatial distribution of galaxies. For three-dimensional distributions it has been found that $\gamma=3/2$ and $\rho_0=9\sigma_{LOS}^2/4\pi Gr_c^2$, while for observed two-dimensional (or projected) distributions $\gamma=1$ and $\rho_0=9\sigma_{LOS}^2/2\pi Gr_c$ \citep[see,][]{ro1972,sc2015}, where $G$ is the gravitational constant. 
Thus, {here we consider} the projection (on the RA-Dec plane) {of a cluster in equilibrium to have} a King-type radial distribution of galaxies {described by the} (1D) radial-PDF of the form
\begin{equation}\label{f_r}
{f}^{\mathrm{eq}}_r(r)=\frac{6r}{\pi r'_c R_{\mathrm{vir}}}\left[1+\left(\frac{r}{r'_c} \right)^2 \right]^{-1}, 
\end{equation}
in which the count of galaxies must be performed in bins of length $\Delta r$ 
%from 0 to $R_{\mathrm{vir}}$ 
instead of rings of area $\Delta A=2\pi r\Delta r$, that is, counting the number of galaxies located between $r$ and $r+\Delta r$ for each distance from the cluster center. 
Integrating (\ref{f_r}) from $0$ to $R_{\mathrm{vir}}$, the normalization condition to find the `relaxed' core radius ($r'_c$) in virial equilibrium is
\begin{equation}\label{nc}
\ln{\left(1+\frac{R_{\mathrm{vir}}^2}{r_c^{'2}}\right)}=\frac{\pi R_{\mathrm{vir}}}{3r'_c}.
\end{equation}

For the azimuthal distribution of galaxies {on the RA-Dec projected plane of a regular cluster we expect} a continuous uniform azimuthal-PDF of the form 
\begin{equation}\label{f_a}
{f}^{\mathrm{eq}}_{\theta}(\theta)= \left\{ \begin{array}{ll}
             1/360, & \mathrm{for} \,\ \theta\in [0, 360] \\
             0,     & \mathrm{otherwise}, \\
             \end{array}
   \right.
\end{equation}
since, under the assumed equilibrium model, the probability of finding galaxies in any direction of the plane must be the same if there are no deformities (flattening or elongation) {in the cluster morphology} and no substructures when counting galaxies in slices of width $\Delta \theta$ at fixed radius. 

%%%%%%%%%%%%%%%%%%%%%%%%%%%%%%%%%%%%%%%%%%%%%%%%%%%%%%%
%%%%%%%%%%%%%%%%%%%%%%%%%%%%%%%%%%%%%%%%%%%%%%%%%%%%%%%

Finally, for the {(3D) galaxy velocities inside clusters} in equilibrium we expect a {quasi-Maxwellian distribution \citep[isotropic, \textit{e.g.},][]{sa1988,sam2014},} {so that the LOS component of velocities have a} normal distribution described by the redshift-PDF 
\begin{equation}\label{f_z} 
{f}^{\mathrm{eq}}_z(z)=\frac{c}{\sigma_{LOS}\sqrt{2\pi}}\exp{\left\lbrace -\frac{c^2}{2}\left(\frac{z-\bar{z}}{\sigma_{LOS}} \right)^2 \right\rbrace},
\end{equation} %=\frac{dP_z}{dz}
where $c$ is the speed of light and $c\bar{z}$ is the mean LOS velocity of the cluster. 

{The construction of the reference PDFs associated to the data is done in three steps. 
%the specific characteristics of each observed cluster (\textit{e.g.}, number of observed galaxies, virial radius and velocity dispersion), 
First,} we determine numerically the relaxed core radius $r'_c$ (see, Column 5 of Table \ref{tab:prop} for observational sample), {taking into account} the normalization condition (\ref{nc}), {to be used in} ${f}^{\mathrm{eq}}_r$. {For} $R_{\mathrm{vir}}$, {also used in ${f}^{\mathrm{eq}}_r$, and for} $c\bar{z}$ and $\sigma_{LOS}$, {the last ones for} ${f}^{\mathrm{eq}}_z$, {we take the previously calculated parameters (see, Columns 10, 4 and 8, respectively, of Table \ref{tab:init} for observational sample)}.
The $f^{\mathrm{eq}}_{\theta}$ distribution is trivial and does not require observational parameters of clusters.

Next, for each cluster we construct its corresponding relaxed mock cluster, which have as many particles as observed galaxies but distributed according to (or following) the corresponding $f^{\mathrm{eq}}_r$, $f^{\mathrm{eq}}_{\theta}$ and $f^{\mathrm{eq}}_z$ 
{equilibrium} 
%references 
PDFs. 

%%%%%%%%%%%%%%%%%%%%%%%%%%%%%%%%%%%%%%%%%%%%%%%%%%%%%
%%%%%%%%%%%%%%%%%%%%%%%%%%%%%%%%%%%%%%%%%%%%%%%%%%%%%

{The third step is done by fitting the tuned} 
%previously expected 
%chosen 
equilibrium models to the particle distributions of the mock clusters in the radial, azimuthal, and redshift components, using bin widths and smoothing {levels} equal to those used in the 
%empirical 
\emph{{observed} PDFs}.
{With this, we are able} to construct the \emph{relaxed PDFs}, $\tilde{f}_r$, $\tilde{f}_{\theta}$ and $\tilde{f}_z$, {\textit{i.e.}, the ones expected in the dynamical relaxation state}. 
Fig. \ref{f:pdf} shows an example of the \emph{observed} and \emph{relaxed PDFs} for the A2199 cluster. 

{Now,} the relaxation probability {of a galaxy system} can be defined as a distance between the \emph{observed PDFs} and the \emph{relaxed} ones, \textit{i.e.}, between the current dynamical state of the galaxy ensemble {(characterized by $\bar{f}_r$, $\bar{f}_z$ and $\bar{f}_\theta$)} and its most probable equilibrium state {(characterized by $\tilde{f}_r$, $\tilde{f}_\theta$ and $\tilde{f}_z$)}, in the probability space. 
For this, we use the Hellinger distance \citep{he1909}, a metric used to quantify the similarity between two distributions in the same probability space so that, if $f_1$ and $f_2$ represent two PDFs for the same variable, the Hellinger distance between them is defined as
\begin{equation}\label{H}
H(f_1,f_2)\equiv\left[\frac{1}{2}\int\left(\sqrt{f_1(x)}-\sqrt{f_2(x)}\right)^2 dx\right]^{1/2},
\end{equation}
where the integration must be carried out over the domain of the functions, and the property $0\leq H(f_1,f_2)\leq 1$ allows us to define the relaxation probability, $\mathcal{P}_{\mathrm{relax}}\equiv 1-H$, of a galaxy ensemble. 
{Thus, the close the $\bar{f}_i$ and $\tilde{f}_i$ functions are to each other, the more relaxed a galaxy system can be considered. For a system close to equilibrium $H\longrightarrow 0$ and $\mathcal{P}_{\mathrm{relax}} \longrightarrow 1$.} 

{Under the same assumption of statistical independence between the variables $r$, $\theta$ and $z$ used for the observed joint PDFs, the relaxed joint PDFs were then defined as $\tilde{f}_{r\theta z}(r,\theta,z)=\tilde{f}_r(r)\tilde{f}_{\theta}(\theta)\tilde{f}_z(z)$. For calculations of $H$, we take $f_1=\bar{f}_{r\theta z}$ and $f_2=\tilde{f}_{r\theta z}$ in (\ref{H}) for each galaxy cluster.} 
The calculated values for relaxation probability, $\mathcal{P}_\mathrm{relax}$, for the observed clusters are shown in Column 6 of Table \ref{tab:prop}. 

\begin{figure}[!tb]
  \begin{center}
   %<left> <lower> <right> <upper>
  \includegraphics[trim={1.3cm 0cm 2cm 0cm},clip,width=\columnwidth]{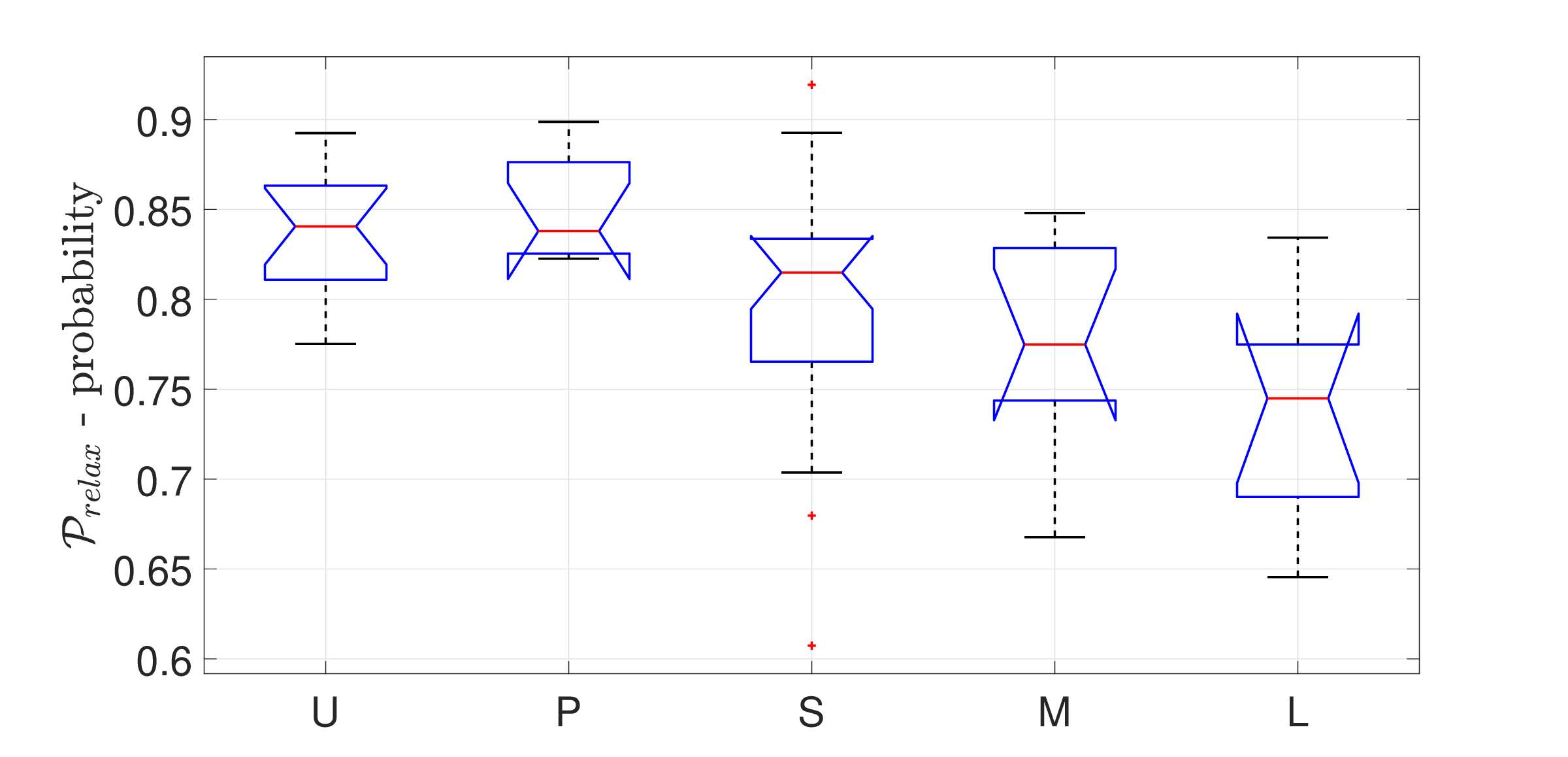}\\
  \includegraphics[trim={1.4cm 0cm 2cm 0cm},clip,width=\columnwidth]{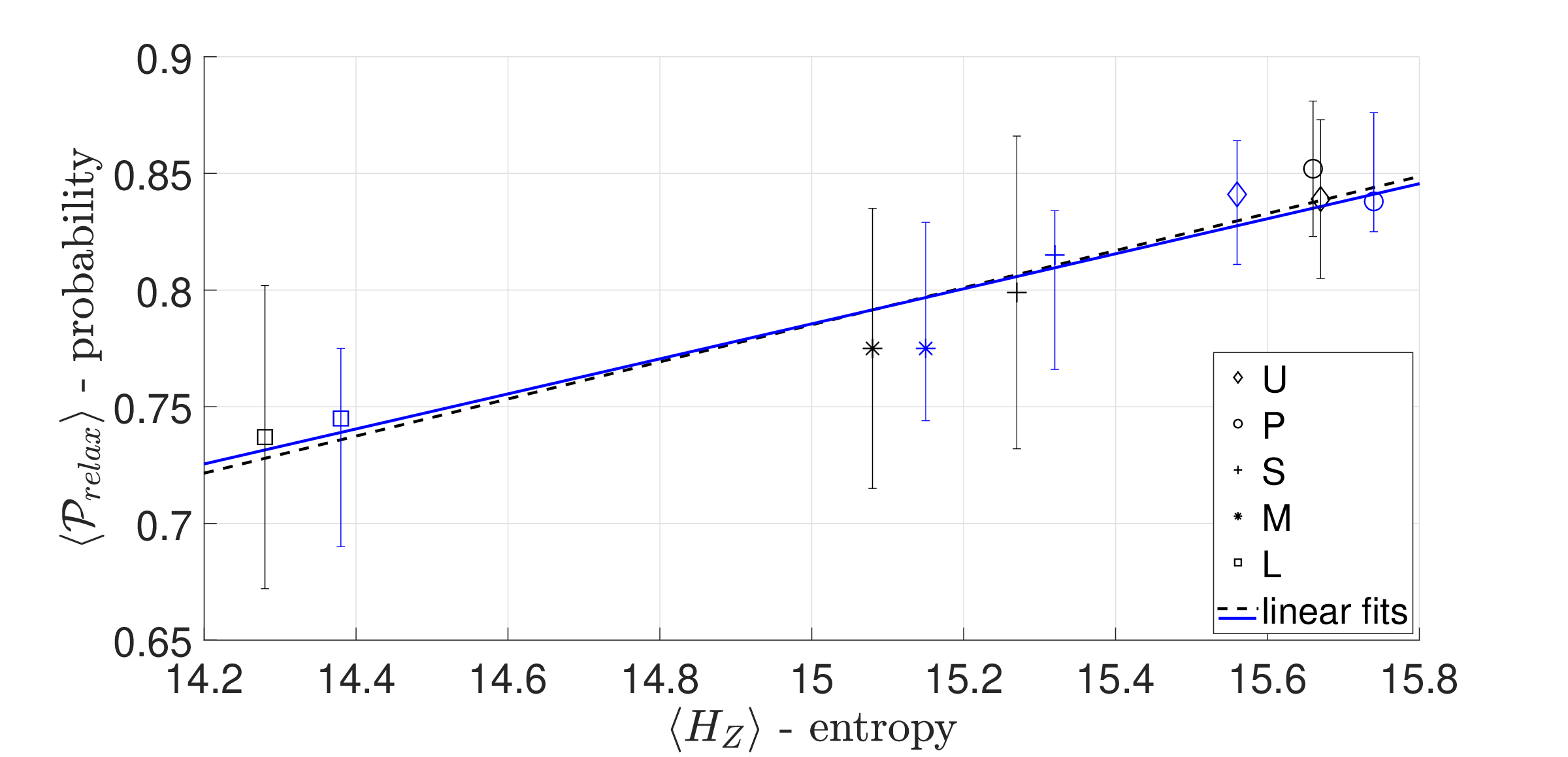}\\
\caption{\textit{Top panel}: Boxplots of $\mathcal{P}_\mathrm{relax}$ values for the five assembly {states} of clusters {from} \citet{ca2021}. {\textit{Bottom panel}: Scatter plot of $\left\langle{\mathcal{P}}_\mathrm{relax}\right\rangle$ \textit{vs.} $\left\langle {H}_Z \right\rangle$ made with mean (black markers) an median (blue markers) $H_Z$ and $\mathcal{P}_\mathrm{relax}$ values presented in Table \ref{t:m_val}. The symbols of the points represents the U-P-S-M-L assembly level classes. The black dashed and blue solid lines represents the linear fits to the mean and median data, with $\mathcal{R}^2$ of 0.964 and 0.950, respectively.}}
 \label{f:correl3}
  \end{center}
\end{figure}

{In addition, the top panel of Fig. \ref{f:correl3} shows the distribution of $\mathcal{P}_{\mathrm{relax}}$ values with respect to the assembly state classes.} Like what happened with the $H_Z$-entropy, there is a clear 
%statistical 
correlation between the relaxation probability and the level of gravitational assembly of a system: the sequence U-P-S-M-L considered here goes from the most likely relaxed to the least relaxed systems with respect to the chosen equilibrium model. {The triple correlation between {dynamical} entropy, the relaxation probability and the assembly level of clusters can be seen in the bottom panel of Fig. \ref{f:correl3}, a scatterplot of $\left\langle{\mathcal{P}}_{\mathrm{relax}}\right\rangle$ \textit{vs.} $\left\langle{H}_Z\right\rangle$ built with the mean and median values (see, Table \ref{t:m_val}) obtained for $H_z$ and $\mathcal{P}_{\mathrm{relax}}$ in each assembly classification.}\\

%%%%%%%%%%%%%%%%%%%%%%%%%%%%%%%%%%%%%%%%%%%%%%%%%%%%%%%
%%%%%%%%%%%%%%%%%%%%%%%%%%%%%%%%%%%%%%%%%%%%%%%%%%%%%%%%

%%%%%%%%%%%%%%%%%%%%%%%%%%%%%%%%%%%%%%%%%%%%%%%%%%%%%%%%
\begin{table*}[!ht]%
\caption{Mean (with standard deviation) and median (with the first-25\% and third-75\% quartiles) values for $H_Z$ and $\mathcal{P}_\mathrm{relax}$ in each of the assembly state classes.}
\label{t:m_val}
\centering %
\resizebox{9cm}{!}{
\begin{tabular}{c cc cc}
\hline\hline
\multicolumn{1}{c}{Assembly level}   & \multicolumn{2}{c}{Mean $\pm$ std} & \multicolumn{2}{c}{Median$_{-Q1}^{+Q3}$} \\
class&$H_Z$ & $\mathcal{P}_\mathrm{relax}$ & $H_Z$ & $\mathcal{P}_\mathrm{relax}$ \\ \hline
U & $15.67 \pm 0.53$ & $0.839\pm 0.034$ & $15.56_{-0.28}^{+0.55}$ & $0.841_{-0.030}^{+0.023}$ \\

P & $15.66 \pm 0.50$ & $0.852\pm 0.029$ & $15.74_{-0.34}^{+0.35}$ & $0.838_{-0.013}^{+0.038}$ \\

S & $15.27 \pm 0.91$ & $0.799\pm 0.067$ & $15.32_{-0.31}^{+0.62}$ & $0.815_{-0.049}^{+0.019}$ \\

M & $15.08 \pm 0.65$ & $0.775\pm 0.060$ & $15.15_{-0.56}^{+0.41}$ & $0.775_{-0.031}^{+0.054}$\\

L & $14.28 \pm 0.58$ & $0.737\pm 0.065$ & $14.38_{-0.57}^{+0.35}$ & $0.745_{-0.055}^{+0.030}$ \\
\hline
\end{tabular}}
\end{table*}
%%%%%%%%%%%%%%%%%%%%%%%%%%%%%%%%%%%%%%%%%%%%%%%%%%%%%%%%

Finally, we applied a Kolmogorov-Smirnov (KS-) test to evaluate, with a confidence level set at 90\% (or significance level of 0.1), the null hypothesis that the $H_Z$ or $\mathcal{P}_\mathrm{relax}$ values for the galaxy systems classified into two different assembly classes (U, P, S, M and L) come from the same continuous distribution. Table \ref{tab:KS_test} shows the $p$-values, in the range $[0, 1]$, that resulted from the KS-test. The closer a $p$-value is to 1, the more similar are the distributions of $H_Z$ and $\mathcal{P}_\mathrm{relax}$ values in two different assembly ($\cal{A}$) classes. 

%%%%%%%%%%%%%%%%%%%%%%%%%%%%%%%%%%%%%%%%%%%%%%%%%%%%%%%%
\begin{table*}
\centering 
\caption{$p$-values of Kolmogorov-Smirnov test comparing the distributions of $H_Z$-entropy (left values) and relaxation probability $\mathcal{P}_\mathrm{relax}$ (right values) for galaxy clusters in different classifications. The confidence level for the KS-test was 90\%.}
%\resizebox{9.5cm}{!}{
%\hspace*{}
\begin{tabular}{c|cc|cc|cc|cc}
\hline\hline 
  \multicolumn{1}{c}{Class}&  
  \multicolumn{2}{c}{P} &
  \multicolumn{2}{c}{S} &
  \multicolumn{2}{c}{M} &
  \multicolumn{2}{c}{L} \\
\hline
 U   & 0.958 & 0.500 & 0.417 & 0.024 & 0.065 & 0.024 & 2.003e-04 & 7.725e-04 \\ 
 P   &      -&-      & 0.205 & 0.006 & 0.203 & 0.037 & 7.652e-04 & 9.766e-04  \\
 S   &      -&-      &      -&-      & 0.509 & 0.714 & 7.044e-04 & 0.019      \\
 M   &      -&-      &      -&-      &      -&-      & 0.148     & 0.243      \\
\hline 
\end{tabular}%} 
\label{tab:KS_test}
\end{table*}
%%%%%%%%%%%%%%%%%%%%%%%%%%%%%%%%%%%%%%%%%%%%%%%%%%%%%%%

% Subsection 6.2
%%%%%%%%%%%%%%%%%%%%%%%%%%%%%%%%%%%%%%%%%%%%%%%%%%%%%%%%%
\subsection{{Relation between $H_Z$ and the cluster concentration indices}}\label{subsec:res4}

{Apart from the discrete characterization of the assembly state used in the previous sections, one can also probe some continuous parameters, estimated from optical or X-ray data  \citep[\textit{e.g.},][and references therein]{car1996,gi2001,zh2011,ca2021},  that are expected to correlate with the evolutionary state of the galaxy system.} 
{Here we present, for instance, the concentration indices of a cluster, both from optical galaxy distributions and X-ray from ICM: more relaxed clusters tend to be denser at their centers and have higher values for these indices.} 

{For the optical data, we use the Maximum Likelihood Estimation method (MLE) to determined the optimal concentration indices for the King and NFW radial profiles fitted to the observed projected distributions of galaxies in the clusters of the \emph{Top70} sample.
%, in order to ascertain their respective concentration indices. 
The King profile employed for the fitting is the one described by equation (\ref{K_3d}), while the NFW profile takes the form 
\begin{equation}
\rho(r)=\rho_0\left[\left(\frac{r}{r_s} \right)\left(1+\frac{r}{r_s} \right)^2 \right]^{-\gamma},
\end{equation}
where $r_s$ is the scale radius of the galaxy cluster. The  concentration indices of each cluster are related to its respective virial and characteristic radii ($r_c$ or $r_s$ obtained by MLE) in the form $c_\mathrm{K}=R_\mathrm{vir}/r_c$ and $c_\mathrm{NFW}=R_\mathrm{vir}/r_s$ \citep[\textit{e.g.}][]{bt2008}, and the obtained values are compiled in Columns 7 and 8 of Table \ref{tab:prop}. 
%On the other hand, we 
We also probed the ratio $c_{500}=R_\mathrm{vir}/r_{500}$ (using the available X-ray data in Table \ref{tab:init}), the obtained values shown in Column 9 of the Table \ref{tab:prop}}.

%%%%%%%%%%%%%%%%%%%%%%%%%%%%%%%%%%%%%%%%%%%%%%%%%%%%
\begin{figure}[t]
  \centering
  %<left> <lower> <right> <upper>
    \includegraphics[trim={1cm 0cm 3cm 0cm},clip,width=\columnwidth]{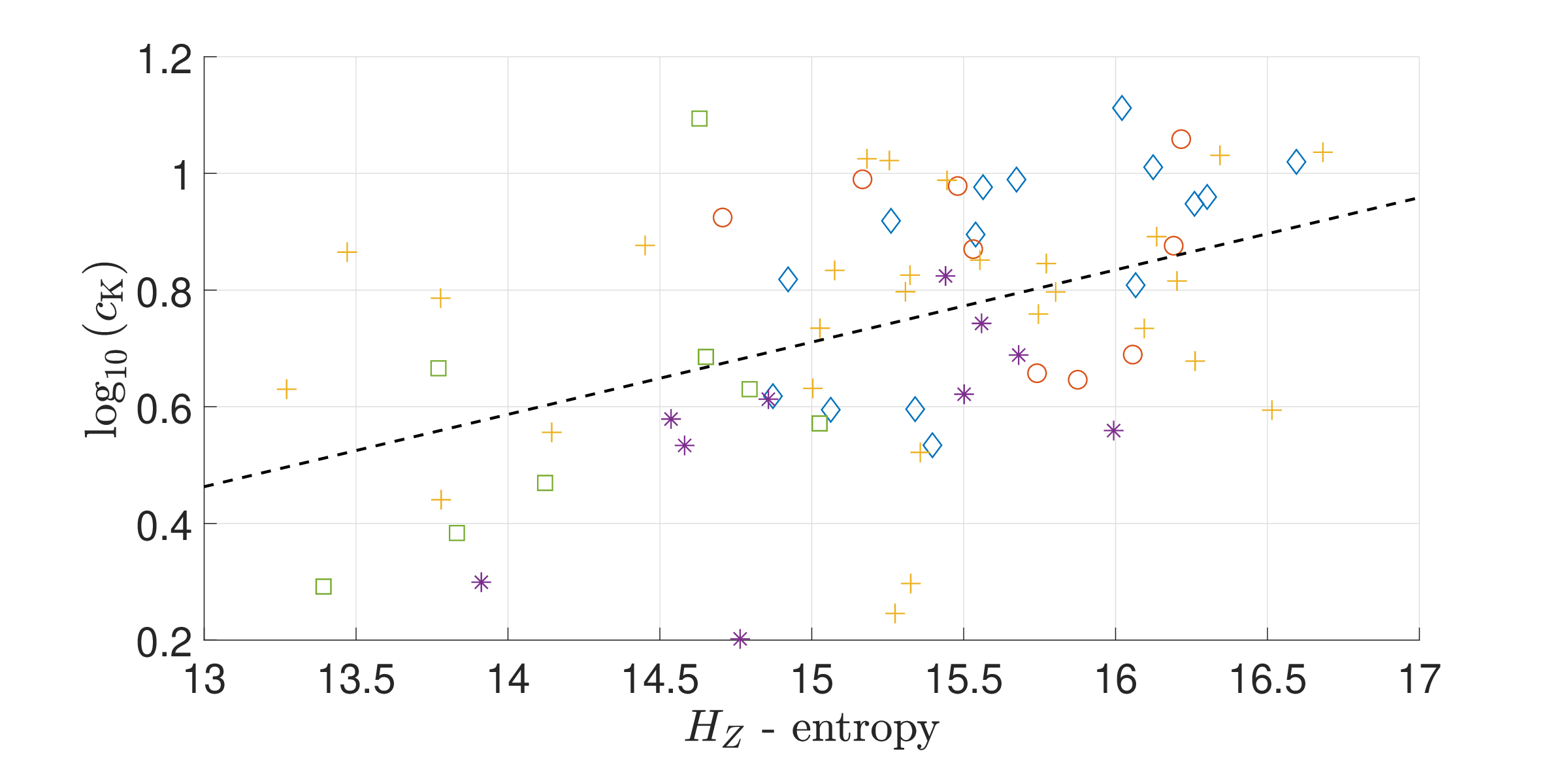}\\
    \includegraphics[trim={1cm 0cm 3cm 0cm},clip,width=\columnwidth]{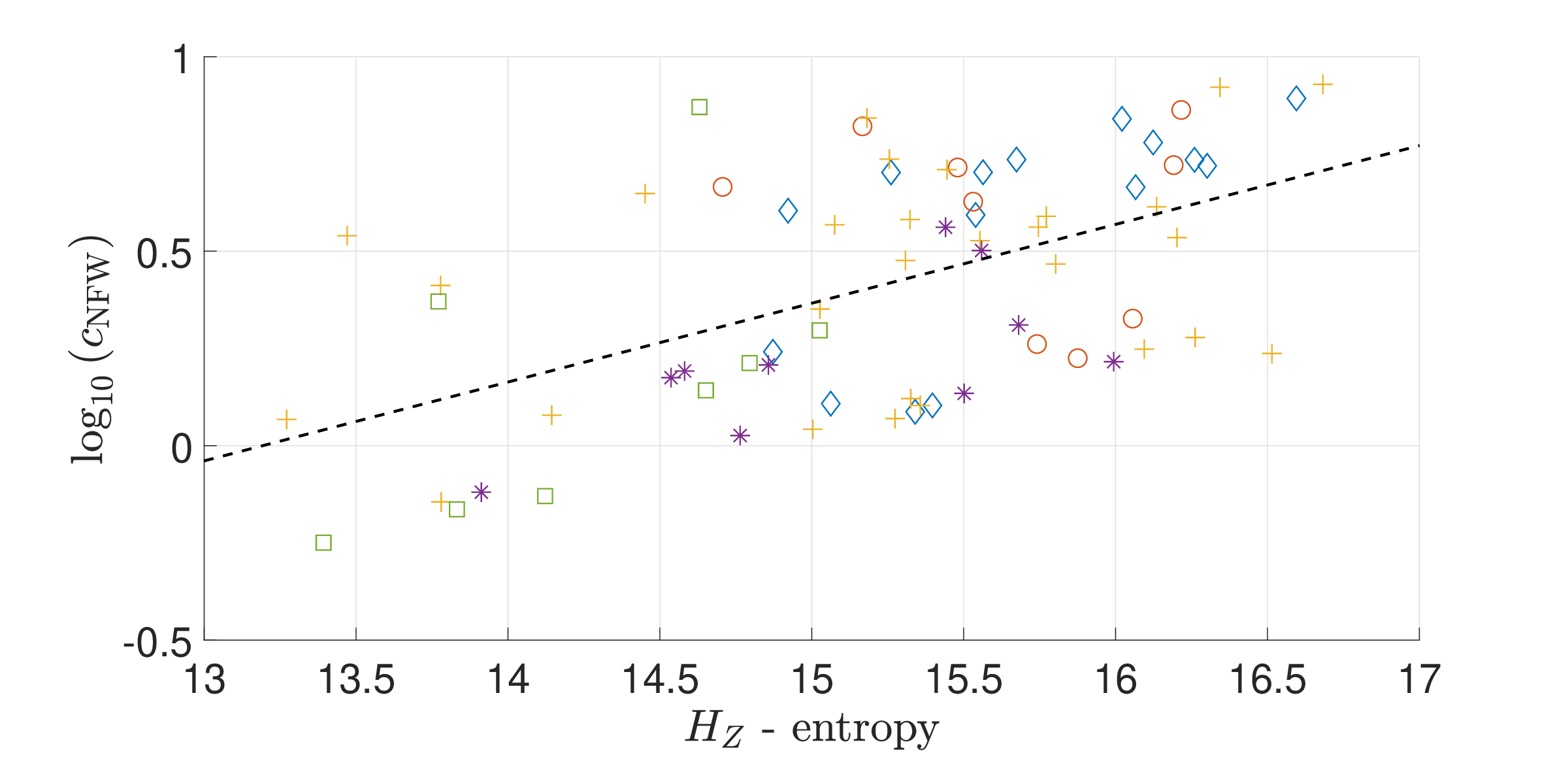}\\
    \includegraphics[trim={1cm 0cm 3cm 0cm},clip,width=\columnwidth]{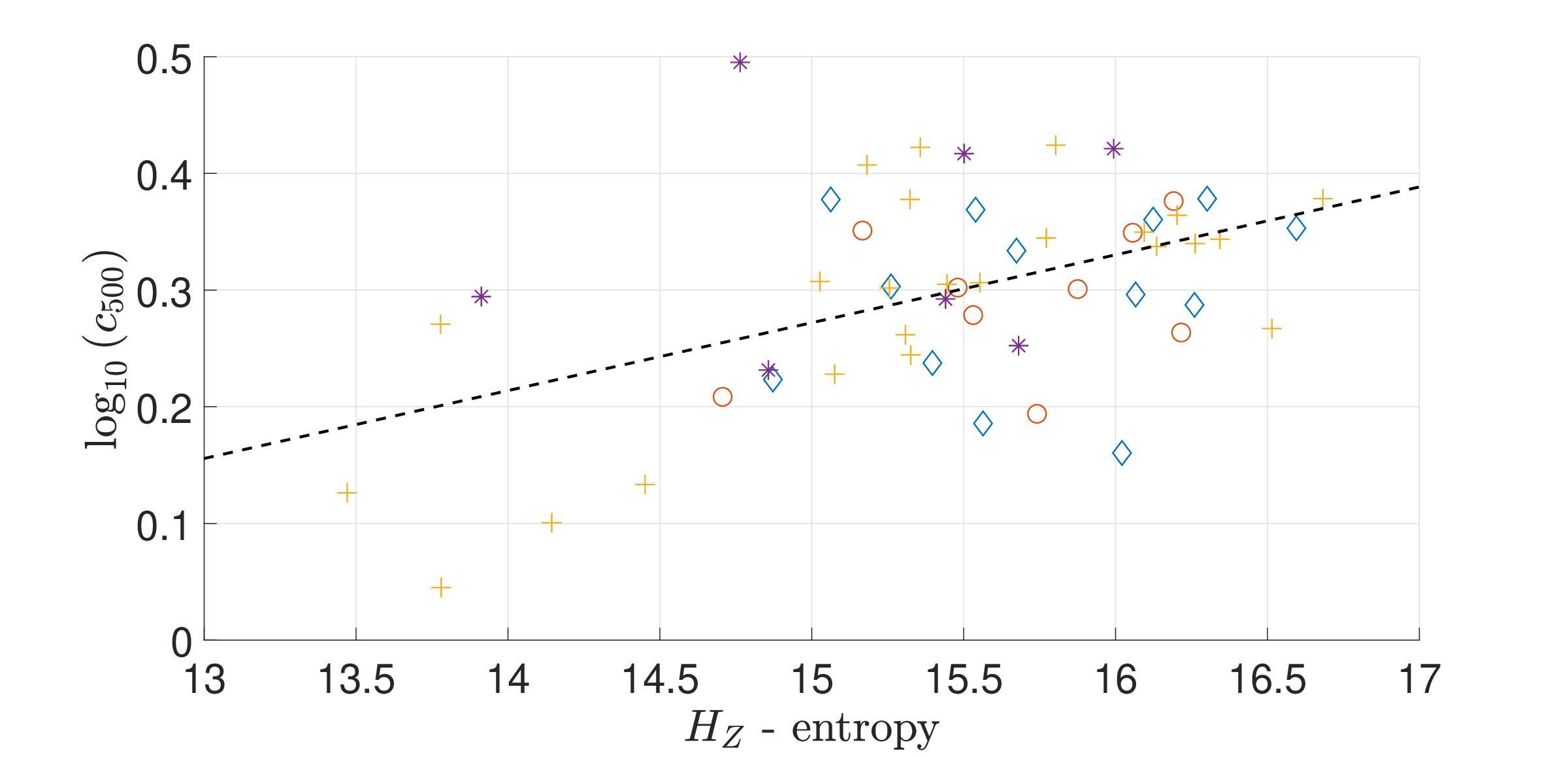}
 \caption{{Scatterplots of $H_Z$-entropy \textit{vs.} concentration indices ($c_\mathrm{K}$, $c_\mathrm{NFW}$, and $c_{500}$) for the \emph{Top70} cluster sample. The symbols used for the markers have the same meaning as in Fig. \ref{f:HzHs}. The dashed line represents the best linear fit to the data, with $\mathcal{R}^2$ of 0.203, 0.264 and 0.215, respectively from top to bottom.}}
 \label{f:conc}
\end{figure}
%%%%%%%%%%%%%%%%%%%%%%%%%%%%%%%%%%%%%%%%%%%%%%%%%%%%%

{Figure \ref{f:conc} shows the relationship between the $H_Z$-entropy and the estimated ---cluster--- concentration indices %for the clusters in
for the \emph{Top70} sample. The Pearson correlation coefficients between the $H_Z$-entropy values and those corresponding to the indices $c_\mathrm{K}$, $c_\mathrm{NFW}$ and $c_{500}$ are 0.448, 0.512, 0.429, respectively. 
However, despite the not strong statistical significance and the remarkable dispersion, an evident growing trend can be seen in the concentration indices with the growth of the $H_Z$-entropy in the systems. This tendency is also subtly manifested in the assemblage classes (represented by the marker symbols in Figure \ref{f:conc}): %such that, statistically speaking, 
the U and P clusters tend to have higher concentration indices, while the L clusters tend to present lower values of these. S and M clusters appear with a more extended range of concentration indices.}  
{It is interesting to note that if we leave only U and P clusters in the above relation, for example for $c_{500}$, the correlation increases from 0.429 to 0.603 ---these are the typical clusters that appear in studies based on X-rays emission of ICM.}

{We have also probed other continuous dynamical parameters, obtained from X-rays emission of ICM, associated to the evolutionary state of galaxy systems: the concentration $c$ and the luminosity concentration $c_L$ \citep[resepctively from,][]{par2015,zh2017}, the Gini and $M_{20}$ morphological coefficients \citep[from][]{par2015,lov2017}, the substructure level $S_C$ \citep[from][]{as2012}, and the $C_\mathrm{SB}$ and $C_\mathrm{SB4}$ concentration parameters \citep[from][]{as2017}. Although the statistics are usually poor, and the aspects captured from X-ray data are not necessarily similar to the ones captured by optical data (which include $H_Z$), the results are all similar to the ones showed for $c_{500}$, for example increasing the correlation when only U and P clusters are considered.}

{It is remarkable that the only parameter that correlates very well with $H_Z$ is $r'_c$ (with a Pearson coefficient of 0.972). The theory states that the natural tendency of a gravitational system towards its evolution is establishing a structure core-halo. This is exactly what the correlation of $H_Z$ and $r'_c$ may be suggesting: the core radius grows with the entropy of the cluster.}

{We consider that $H_Z$ is an efficient parameter for estimating the relaxation/dynamical state of galaxy clusters, maybe better than the the others we have compared in this not complete exercise, because it can capture more important aspects of such evolutionary snapshot.
%than the other parameters we have compared in this not complete exercise. 
However, it is important to note that no parameter can, alone, account for all the evolutionary aspects such as interaction with the cosmological environment, collapse and mass growth, galaxy formation, AGN activity, feedback processes, among others, making paramount to consider different observations and analyses for constructing the most complete picture as possible.}

%%%%%%%%%%%%%%%%%%%%%%%%%%%%%%%%%%%%%%%%%%%%%%%%%%%%%%%%%
% Section 7
%%%%%%%%%%%%%%%%%%%%%%%%%%%%%%%%%%%%%%%%%%%%%%%%%%%%%%%%%
\section{Discussion and Conclusions}
\label{sec:disc}

In this work we {have proposed an} entropy estimator, {$H_Z$, {which can be calculated} from global dynamical parameters (virial mass, projected virial radius and velocity dispersion),} to characterize the dynamical state of galaxy systems. 
Initially, a slight modification of the standard $T\,ds$ Gibbs relation was carried out to include the peculiar behavior of self-gravitating systems and, as a result, an expression was obtained for the entropy component $s$ related only to the galaxy ensemble, which is a function of the internal kinetic energy and the volume of the systems.

%%%%%%%%%%%%%%%%%%%%%%%%%%%%%%%%%%%%%%%%%%%%%%%%%%%%%%%%%
% Discussion about $H_Z%
{The direct association of} the $H_Z$-entropy estimator {to} the dynamical state of the {galaxy} systems
%, we resorted to 
{comes from} the second law of thermodynamics,  according to which entropy increases as the systems advance towards more stable states, reaching a local maximum in virial equilibrium. This does not assert that the galaxy system reaches a stable state of equilibrium, which would be determined by a global maximum of entropy, but rather a state of greater stability than {its} previous evolutionary configurations, 
%of assembly and substructuring, 
the so-called dynamical relaxation. 
{Thus}, although the virial theorem is fulfilled in relaxed states with local entropy maximum, these are \emph{metastable equilibria} that can be broken {if the galaxy system} significantly {interacts} with its environment, such as through {accretions and} mergers with other systems. {However, if} the system is isolated (or its interaction with the environment is negligible), its entropy value remains in the vicinity of the maximum and its behavior in such state will be indistinguishable from dynamical relaxation.

%%%%%%%%%%%%%%%%%%%%%%%%%%%%%%%%%%%%%%%%%%%%%%%%%%%%%%%%%
% Tests
{In order to evaluate the power of this entropy to represent the evolutionary state of a galaxy system, we correlated it to four independent estimators: two observational and two statistical. The two observational come from analyses of the internal structure of real clusters, particularly the discrete gravitational assembly state (obtained from optical data) and three continuous concentration indices (both from optical and X-ray data),} applied {to} a sample of 70 {well spectroscopically-sampled} nearby galaxy clusters. 
{The statistical estimators comprehend the Shannon (information) entropy and the relaxation probability, calculated for the same sample of observed clusters and for a sample of 248 halos (simulated clusters) from IllustrisTNG.}

%%%%%%%%%%%%%%%%%%%%%%%%%%%%%%%%%%%%%%%%%%%%%%%%%%%%%%%%%
% $\mathcal{A}$

{The first} striking result {of our analysis is} that the $H_Z$-entropy correlates very well with the {gravitational} assembly state of the clusters obtained from observational {optical data.} 
%(see, bottom panel of Figure \ref{f:HzHs}). 
The $H_Z$-entropy increases with the decrease in the level of substructuring, which is interpreted as the less relaxed a cluster is, more information is lost when treating it as a virialized system. 
Specifically, both U and P clusters show the highest entropies, while L present the lowest values for this property. Since P clusters {are massive, while possessing} only low significance substructures (low mass accretions), they resemble the unimodal U clusters. On their turn, L clusters are, as pointed by \citet{ca2021}, {the `less evolved', in the sense that they are} the poorest, less massive and have not suffered significant merging processes yet.
%the poorest clusters of the sample and, since they are  
%, the ones with lower entropies.
S and M clusters present a large range in entropies. This happens because both less or more massive clusters (and both less and more evolved ones) can suffer new mergings or accretion at any time.  

{The $H_Z$-entropy estimator was derived based on specific variables (\textit{e.g.}, per mass unit), which allows comparisons between galaxy systems of different masses ($\mathcal{M}_{\mathrm{vir}}$) and sizes ($R_{\mathrm{vir}}$).
%(since $R_{\mathrm{vir}}\propto M_{\mathrm{vir}}$ as shown by (\ref{R_v})). 
{Moreover}, equation (\ref{s2}) shows a clear functional dependence of the entropy on the galaxy velocity dispersion (\textit{i.e.}, on the specific internal kinetic energy $\kappa=(\alpha/2)\sigma_{LOS}^2$ of the galaxy ensemble). This implies that although groups and poor clusters can reach virial equilibrium like rich clusters, the relaxation state of the latter will be characterized by higher entropy values given their higher velocity dispersions.} 
{It is possible to verify that, by calculating $H_Z$ without taking into account the term that depends on $\sigma_{LOS}$ in (\ref{s2}), then the entropy values of all sampled clusters are very closely distributed around a central value $\langle H_Z\rangle$ that depends on the units chosen for $\mathcal{M}_\mathrm{vir}$ and $R_\mathrm{vir}$.}

%%%%%%%%%%%%%%%%%%%%%%%%%%%%%%%%%%%%%%%%%%%%%%%%%%%%%
\begin{figure*}
  \begin{center}
       \includegraphics[trim={0cm 0 0cm 0},clip,scale=0.6]{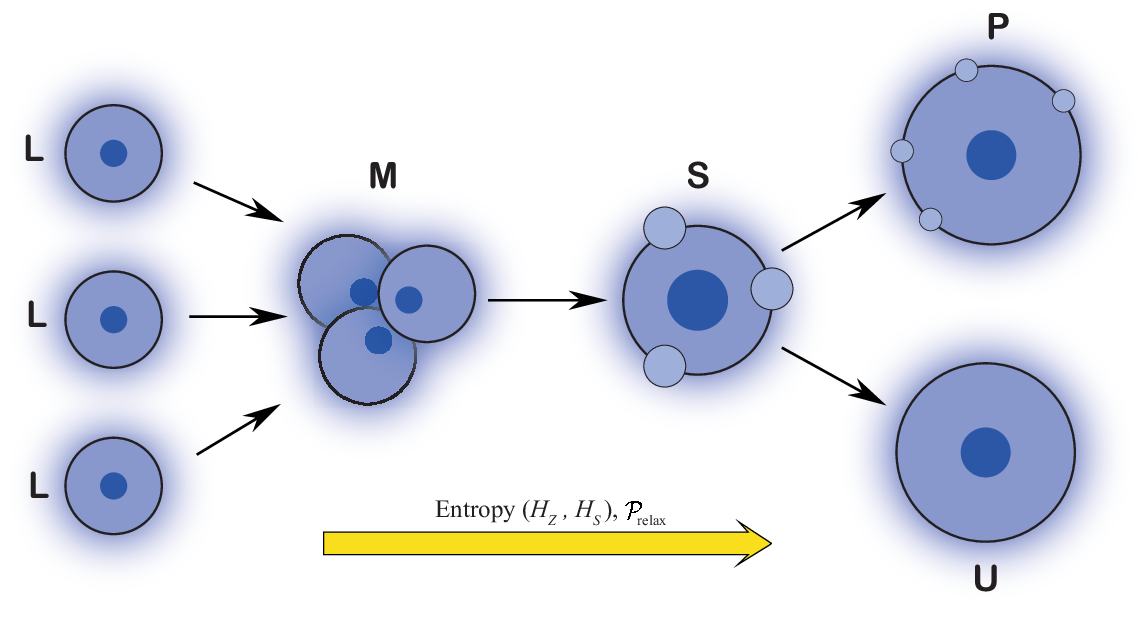}%upsml
\caption{Diagram of a possible evolutionary line between systems in the different assembly levels (U-P-S-M-L). The entropy levels of the galaxy systems, obtained using the $H_Z$ and $H_S$ estimators, increase as they progress towards more dynamically relaxed states (with a higher probability of relaxation, $\mathcal{P}_\mathrm{relax}$), \textit{i.e.}, states with more homogeneous and less substructured galaxy distributions (or where the substructures are insignificant for the dynamics of the systems).}
 \label{f:ev}
  \end{center}
\end{figure*}
%%%%%%%%%%%%%%%%%%%%%%%%%%%%%%%%%%%%%%%%%%%%%%%%%%%

A possible evolutionary line between systems at different assembly levels has been schematized in Figure \ref{f:ev}. Multimodal (M) systems could be formed by the merger of two or more low-mass unimodal (L) systems. These mergers increase the entropy level of the systems due to the increase in mass and number of particles (\textit{i.e.}, galaxies). As  gravity further assembles the fused parts, a main structure is formed in the clusters, and, when  still dynamically significant substructures remain, we have substructured systems (S). In this process, the entropy increases because the main structure, which is larger than the rest of the substructures, begins a virialization process and dominates the dynamics of the cluster. The substructures are special configurations in the distribution of galaxies, macroscopic movements that can slightly affect the dynamics of the system. However, these are little by little accreted and dissolved by the main structure, until they become less significant and the clusters evolve marginally towards the more assembled states primaries (P) and unimodals (U), where the entropy is greater given the large homogeneity in the distribution of galaxies, large velocity dispersion and the proximity to virialization. 

{It is important to highlight that, despite the notable tendency of $H_Z$-entropy to increase with the L$\rightarrow$M$\rightarrow$S$\rightarrow$P/U sequence of assembly states, an individual cluster could not be assigned to a specific $\cal{A}$ class only knowing its $H_Z$-entropy value, since this classification requires the knowledge of more observational (optical and X-ray) features of the cluster. However, the $H_Z$-entropy values allow us to know statistically which clusters are likely to be more relaxed ---and probably with a higher assembly level--- than others when working with a large sample of clusters, \textit{e.g.}, as in the case of the \emph{Top70} or TNG300 sample. 
In addition, if by qualitative methods two clusters appear similar, their $H_Z$ values can help to discriminate if they are at the same evolutionary state or if one of them is more advanced (or delayed) than the other, like is the case of L and U clusters.}

%%%%%%%%%%%%%%%%%%%%%%%%%%%%%%%%%%%%%%%%%%%%%%%%%%%%%%%%%

{Resuming the question about unimodal clusters, we have also verified that L clusters lack the core-halo spatial configuration,  
%and rarely have a central core, 
unlike U clusters that exhibit a clear central concentration of galaxies. In fact, as can be seen from the galaxy concentration indices,
%(see Table \ref{tab:prop} and Fig. \ref{f:correl3}), 
U clusters generally achieve the highest concentration indices while L clusters have the lowest. Thus, the Hellinger distance between the respective \emph{observed} and \emph{relaxed PDFs} of a cluster become larger when the latter lack a central concentration.}

%%%%%%%%%%%%%%%%%%%%%%%%%%%%%%%%%%%%%%%%%%%%%%%%%%%%%%%%%
% $H_S$
{Concerning the statistical estimators we used to test $H_Z$-entropy, we} found that the Shannon ---or information--- entropy $H_S$ 
presents a remarkable (almost linear) correlation with it.

Information entropy does not necessarily have a simple correspondence with physical entropy, but we can provide a possible explanation for the correlation between the $H_Z$ and $H_S$ entropies as follows. Given that the cylinder of $x$-points of observational data constitutes a projected phase-space (\textit{e.g.}, it contains 2D-position and 1D-velocity coordinates of particles in a system that evolves with time) for each cluster, $\bar{f}_{r\theta z}$ is like a `coarse-grained distribution function' obtained for a fixed time ---the observing time--- in this phase-space. 
Thus, the Shannon entropy $H_S(\bar{f}_{r\theta z})$ meets in this context the criteria to be a H-function \citep[see,][{do not confuse with the Hellinger distance in (\ref{H})}]{thl1986} and, therefore, it always increases with the evolution of the galaxy ensemble.

On the other hand, the argument $\kappa^{\alpha/2}\upsilon^{-1}$ in (\ref{s}) is equivalent, by analogy with statistical mechanics concepts, to the $\mathcal{Z}$ partition function of the galaxy ensemble. Every $\mathcal{Z}$ function is defined in the phase-space of a system \citep[\textit{e.g.},][]{hi1956,ll1980}, taking higher values in states of ---or close to--- equilibrium, where dynamical relaxation increases the randomness in the particle distribution and offers a greater number of ways in which energy can be distributed inside the system, \textit{i.e.}, in states of higher entropy since $s\propto \mathcal{Z}$. The $\bar{f}_{r\theta z}$ distributions and $\mathcal{Z}$ functions are intrinsically related in the thermodynamic limit of a system \citep[\textit{e.g.},][]{ja1957}, even in self-gravitating ones \citep[\textit{e.g.},][]{cha2003,cha2006}, reason why both $H_Z$ and $H_S$ are related between them and to the dynamical state of the galaxy ensemble, and allow us to measure how close it is to virial equilibrium.

%%%%%%%%%%%%%%%%%%%%%%%%%%%%%%%%%%%%%%%%%%%%%%%%%%%%%%%%%
% $\mathcal{P}_{relax}$

This interpretation is reinforced by the relaxation probability $\mathcal{P}_\mathrm{relax}$ which shows that the closer the galaxy cluster is to the virialization, smaller is the distance between its observed distribution of galaxies and the expected theoretical equilibrium one.

%%%%%%%%%%%%%%%%%%%%%%%%%%%%%%%%%%%%%%%%%%%%%%%%%%%%%%%%%
% Final

The entropy estimations presented here are very promising because, once one has a representative sample of galaxy members for the cluster, the calculation of the input parameters is straight forward. This low-cost analysis is much easier than the broader one presented in \citet{ca2021}, and can also be used as a complementary analysis in the study of the assembly state of the galaxy systems. It is still lacking a deeper analysis of the implications of the results presented here, what we plan to do in a future work. We also intend to extend this analysis to other scales of galaxy clustering, especially in the direction of the evolution of the large scale structures like the superclusters of galaxies.

{Our main conclusions are the following:
\begin{itemize}
    \item The $H_Z$-entropy estimator, which depends solely on observational (optical) parameters, adequately captures the entropy of clusters manifested in the (spatial and velocity) distribution of their member galaxies.
    \item The $H_Z$-entropy is related, through the second law of thermodynamics, with the evolutionary state of galaxy systems. Entropy increases as systems evolve towards more stable states, reaching a local (non-unique) maximum at virial equilibrium.
    \item There is a significant correlation between the $H_Z$-entropy of galaxy systems and their gravitational assembly states \citep[$\cal{A}$, ][]{ca2021}, presenting an entropy growth in the L$\rightarrow$M$\rightarrow$S$\rightarrow$P/U sequence, direction in which the relaxation probability $\mathcal{P}_\mathrm{relax}$ of the clusters also increases.
    \item There is a remarkable (almost linear) correlation between $H_Z$-entropy and the Shannon (information) entropy, $H_S$, reinforcing that the dynamical entropy we propose can capture the increase in disorder and loss of information in the process of virialization.
    \item Clusters with higher velocity dispersions of member galaxies tend to have higher $H_Z$ and $H_S$ entropy values, indicating more random galaxy distributions.
    \item The $H_Z$-entropy estimator shows a great capacity to capture the state of relaxation and evolution of the galaxy systems, maybe larger than the one presented by other conventional continuous parameters used for this purpose.
    \item The $H_Z$-entropy estimator provides valuable information on the dynamical state and assembly levels of galaxy clusters, which may have significant applications in studying the evolution of galaxy systems and understanding their dynamics.
\end{itemize}}

\section*{Acknowledgments}
The authors JMZ, CAC and EGM thank financial support from \textit{Universidad de Guanajuato} (DAIP), \textit{Convocatoria Institucional de Investigaci\'on Cient\'ifica}, projects 096/2021 and 162/2022.
JMZ acknowledges funding from CONACyT PhD grant and appreciates the manuscript revision by Manuela Paz. APG acknowledges support from \textit{Departamento de Matem\'aticas, Universidad del Cauca}. The authors are grateful to Dr. Roger Coziol and Dr. Francisco Escamilla for the discussions that helped to improve the manuscript.

%%%%%%%%%%%%%%%%%%%%%%%%%%%%%%%%%%%%%%%%%%%%%%%%%%%%%%%%
% Bibliography
%%%%%%%%%%%%%%%%%%%%%%%%%%%%%%%%%%%%%%%%%%%%%%%%%%%%

\end{document}